\documentclass[acmsmall]{acmart}

\AtBeginDocument{%
  \providecommand\BibTeX{{%
    \normalfont B\kern-0.5em{\scshape i\kern-0.25em b}\kern-0.8em\TeX}}}

\setcitestyle{numbers,nosort&compress}
\newcommand{\showDOI}[1]{\unskip}

\setcopyright{acmcopyright}
\acmDOI{10.1145/3527156}
\acmJournal{CSUR}

\usepackage{lineno}
\newcommand{\R}[1]{}
\newcommand{\lr}[1]{}

\usepackage{subcaption}
\usepackage{graphicx}
\usepackage{array,multirow}
\usepackage{amsmath}
\DeclareMathOperator*{\argmin}{argmin}
\usepackage{tablefootnote}
\usepackage{xcolor}
\usepackage{threeparttable}
\usepackage{relsize}
\usepackage{enumitem}
\usepackage{float}
\usepackage{footnote}
\usepackage{threeparttable}

\usepackage{array}
\usepackage{threeparttable}
\usepackage{xcolor}

\usepackage{lscape}

\usepackage{pifont}
\usepackage{newunicodechar}
\newunicodechar{✓}{\ding{51}}
\newunicodechar{✗}{\ding{55}}

\usepackage{makecell}
\usepackage{layout}

\usepackage{xcolor}

\newcommand{\blue}[1]{{\color{black}#1}}

\newcommand{\yellow}[1]{{\color{black}#1}}

\begin{document}

\title{Hardware Approximate Techniques for Deep Neural Network Accelerators: A Survey}

\author{Giorgos~Armeniakos}
\affiliation{%
  \institution{National Technical University of Athens}
  \city{Athens}
  \country{Greece}}
\email{armeniakos@microlab.ntua.gr}

\author{Georgios Zervakis}
\affiliation{%
  \institution{Karlsruhe Institute of Technology}
  \city{Karlsruhe}
  \country{Germany}}
\email{georgios.zervakis@kit.edu}

\author{Dimitrios~Soudris}
\affiliation{%
  \institution{National Technical University of Athens}
  \city{Athens}
  \country{Greece}}
\email{dsoudris@microlab.ntua.gr}

\author{J\"org Henkel}
\affiliation{%
  \institution{Karlsruhe Institute of Technology}
  \city{Karlsruhe}
  \country{Germany}}
\email{henkel@kit.edu}

% !TEX root = ./paper/paper.tex

\begin{abstract}
Deep Neural Networks (DNNs) are very popular because of their high performance in various cognitive tasks in Machine Learning (ML).
Recent advancements in DNNs have brought beyond human accuracy in many tasks, but at the cost of high computational complexity.
To enable efficient execution of DNN inference, more and more research works, therefore, exploit the inherent error resilience of DNNs and employ Approximate Computing (AC) principles to address the elevated energy demands of DNN accelerators.
This article provides a comprehensive survey and analysis of hardware approximation techniques for DNN accelerators.
First, we analyze the state of the art and by identifying approximation families, we cluster the respective works with respect to the approximation type.
Next, we analyze the complexity of the performed evaluations (with respect to the dataset and DNN size) to assess the efficiency, the potential, and limitations of approximate DNN accelerators.
Moreover, a broad discussion is provided, regarding error metrics that are more suitable for designing approximate units for DNN accelerators as well as accuracy recovery approaches that are tailored to DNN inference. 
Finally, we present how Approximate Computing for DNN accelerators can go beyond energy efficiency and address reliability and security issues, as well. 
\end{abstract}

\begin{CCSXML}
<ccs2012>
   <concept>
       <concept_id>10002944.10011122.10002945</concept_id>
       <concept_desc>General and reference~Surveys and overviews</concept_desc>
       <concept_significance>500</concept_significance>
       </concept>
   <concept>
       <concept_id>10010583.10010600.10010615</concept_id>
       <concept_desc>Hardware~Logic circuits</concept_desc>
       <concept_significance>500</concept_significance>
       </concept>
   <concept>
       <concept_id>10010520.10010521</concept_id>
       <concept_desc>Computer systems organization~Architectures</concept_desc>
       <concept_significance>500</concept_significance>
       </concept>
   <concept>
       <concept_id>10010147.10010257.10010293.10010294</concept_id>
       <concept_desc>Computing methodologies~Neural networks</concept_desc>
       <concept_significance>500</concept_significance>
       </concept>
 </ccs2012>
\end{CCSXML}

\ccsdesc[500]{General and reference~Surveys and overviews}
\ccsdesc[500]{Hardware~Logic circuits}
\ccsdesc[500]{Computer systems organization~Architectures}
\ccsdesc[500]{Computing methodologies~Neural networks}

\keywords{Approximate Computing, Arithmetic Circuits, Deep Neural Networks, Error Metrics, Hardware Approximation}
\maketitle

% !TEX root = ./paper.tex
\section{Introduction}\label{sec:introduction}

%\IEEEPARstart{T}{he} quintillion bytes of data that are produced by humans and machines every day, far outpace human’s ability to interpret and make complex decisions.
Advancements in Deep Learning (DL) with Deep Neural Networks (DNNs) delivered \blue{beyond human} levels of accuracy on many AI tasks~\cite{swagath2020}.
\blue{Increasing number} of embedded devices rely on DL and DNNs to deliver sophisticated services such as machine translation~\cite{2018:matchinetranslat}, object detection~\cite{ObjectDetection}, healthcare~\cite{healthcare, healthcare2} etc.
% 2016:machinetranslat, 
% ObjectDetection2
However, these accuracy improvements came at the cost of a vast increase in computational demands, leading to the emerge of customized hardware DNN accelerators\blue{~\cite{google:tpu, swagath2020}}.\label{commentR3C1}
It is noteworthy that recent Convolution Neural Networks (CNNs) require tens of billions of multiply-accumulate (MAC) operations~\cite{swagath2020}.
To satisfy such demands DNN accelerators integrate thousands of MAC units, e.g., Google TPU~\cite{google:tpu} comprises 64K MACs, while Samsung's neural processing unit (NPU) contains 6K MAC units~\cite{Park:ISSCC2021:samsung6knpu}.
This immense number of MAC units combined with high parallelization results in high energy demands.
This problem is intensified, especially when considering the growth of Edge AI that requires \blue{even more} complex neural networks (NNs) to operate on a wide spectrum of energy and resource restricted devices.

\begin{figure}[t]
 \centering
\includegraphics[width=0.73\textwidth]{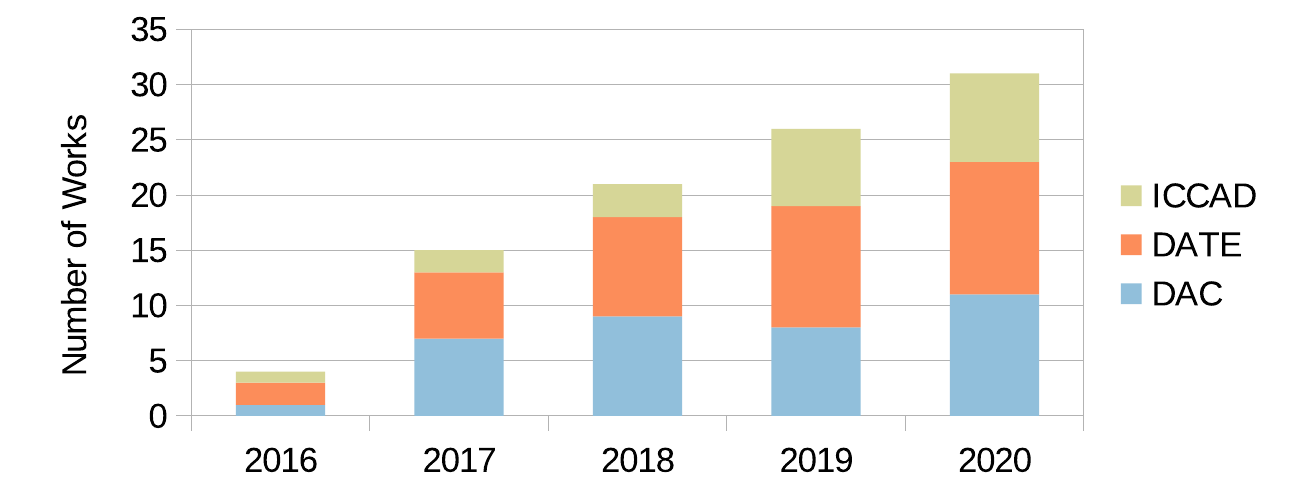}
\centering
\caption{
Number of publications that apply any type of approximation on DNN inference. The past five years and three major design automation conferences are considered.
}
\label{fig:confer}%
\end{figure}

Over the past decade, Approximate Computing (AC)~\cite{Han:ETS2013} established as a new design paradigm for energy efficient circuits.
%\orange{The increasing size of such circuits has led to various exploration and optimization techniques in computer-aided design (CAD)~\cite{mlcad:2021}}.
AC goes beyond typical/emerging design approaches~\cite{mlcad:2021} and exploits the inherent ability of a large number of applications to produce results of acceptable quality, despite some errors (approximations) in their computations.
Leveraging this property, AC approximates the hardware execution of the error resilient computations in a manner that favours performance and energy~\cite{swagath2020}.
% Although AC has been applied with remarkable results in many application domains, such as image processing and statistical machine learning~\cite{Zervakis:ASPDAC2021, swagath2020}, our focus in this article regards the emerging NN workloads.
Driven by the high potential for energy efficiency and exploiting the error tolerance of NNs~\cite{ApproxANN:Date2015, AxNN:Roy2014}, research on approximate NN implementations is rapidly growing over the last years.
Fig.~\ref{fig:confer} is a representative example of this trend.
Fig.~\ref{fig:confer} depicts the number of publications, in three major design automation conferences, that apply approximations in CNN inference.

Considering the high demand for edge AI~\cite{Hao:MDAT2021:enabling}, the billions of mobile devices running DNN inference, and the rapid growth of AI chips\footnote{Google~\cite{google:tpu}, Samsung~\cite{Park:ISSCC2021:samsung6knpu}, Intel~\cite{WechslerHCS2019intel}, IBM~\cite{Agrawal:ISSCC2021:ibmai7nm}, Huawei~\cite{Liao:HCS2019:DaVinci}, Cerebras~\cite{Cerebras:Wafer}, Groq~\cite{Groq:TSP}, Graphcore~\cite{Graphcore:IPU}, Arm~\cite{Arm:Ethos}, NVIDIA~\cite{Nvidia:TensorCore}, etc.},
our focus in this survey is to study, analyze, and elucidate the impact of hardware approximation techniques on the efficiency and accuracy of DNN inference accelerators.
%\blue{DNN accelerators consume most of the total power typically in DRAM~\cite{eden} data movements (dominating between 30\% and 80\% of the system energy~\cite{Yang:ASILOMAR:2017}) and processing elements that usually includes MAC operations.\label{commentR3C3}
\blue{Prior research on DNN accelerators reports that between 30\% to 80\% of the system energy is consumed by DRAM~\cite{eden}
with data movement dominating the energy consumption~\cite{Yang:ASILOMAR:2017}.
% Still, the processing units (e.g., MACs) of DNN accelerators consume significant amount of power (e.g., more
% than 80\% of the total power in~\cite{swagath2020}).
%\yellow{Still, in some cases the processing units (e.g., MACs) of DNN accelerators dominate energy consumption~\cite{swagath2020}.}\label{commentR3C3}
\R{comment_minor_R2C1}\yellow{Still, the processing units (e.g., MACs) of DNN accelerators feature considerable power consumption~\cite{swagath2020,Amrouch2020:NpuThermal}}
Hence, considering high utilization and continuous operation, high energy is also consumed by the processing units that could be prohibitive, for example, in battery power embedded devices~\cite{swagath2020}.
In addition, the very high power consumed by the processing units in a confined area may lead to unsustainable power densities with far reaching impact on the temperature, performance, and reliability of DNN accelerators~\cite{Amrouch2020:NpuThermal}.}
Although several works examine approximate memories for DNNs~\cite{ApproxMemory, Deng2015ReducedMemory, DRAM, eden} such works are out of the scope of our survey which focuses on computational approximation.
Note nevertheless, that compute-based (our survey) and memory-based approximations \blue{are mainly} complementary.
Finally, although approximate computing mainly targets energy efficiency in DNN accelerators (Sections~\ref{sec:techniques}-\ref{sec:evaluation}), several works apply approximations to tackle reliability and security issues (Section~\ref{sec:other}).

The ever-increasing demand for efficient DNN inference as well as the prominent outcomes of AC applications have attracted significant research interest.
As shown in Table~\ref{table:surveys}, several surveys address similar topics with our work.
A survey of approximate arithmetic units (e.g., adders and multipliers) is presented in~\cite{ApproxCircuitsSurvey}.
Nevertheless, in~\cite{ApproxCircuitsSurvey}, only a simple DNN use case example is used as a proof of concept.
On the other hand,~\cite{Reda:AxBook2018} presents a comprehensive study of approximate circuits, discussing also DNN specific approximation techniques.
However, in~\cite{Reda:AxBook2018}, software-based approximation techniques (such as quantization and pruning) are mainly reviewed, while regarding hardware-based approximation, only a limited discussion based on approximate multipliers is included.
In~\cite{swagath2020} and~\cite{Chen:DATE2018} the impact of DNN approximation techniques is reviewed with main focus on software-based approaches.
In~\cite{Chen:JENG2020}, a survey of DNN accelerator architectures is provided while~\cite{Capra:ACCESS2020} reviews hardware and software optimization methods for DNN accelerators.
\blue{Similarly to~\cite{Capra:ACCESS2020},~\cite{Quant_tutorial} and~\cite{Quant_tutorial2} present very comprehensive surveys on software optimizations/approximations and hardware architectures for DNNs.
However, hardware DNN approximations are not the target of~\cite{Chen:JENG2020,Capra:ACCESS2020,Quant_tutorial,Quant_tutorial2}.}\label{commentR1C1}
Finally,~\cite{gholami2021QuantSurvey, liang2021:PrunandQuantSurvey} present a thorough analysis of software based approximation methods such as quantization and pruning, while~\cite{ren2021ArchitectSurvey} provides a comprehensive review of recent NN architectures.
Approximate DNN accelerators are out of the scope of ~\cite{gholami2021QuantSurvey,liang2021:PrunandQuantSurvey,ren2021ArchitectSurvey} %\blue{and although~\cite{Quant_tutorial2} is also a very comprehensive survey that covers meticulously data compression and quantization methods, it still remains singularly focused on DNN compression with almost no mention to hardware-level AC principles.}
\emph{On the other hand, our work surveys the state of the art of approximate DNN accelerators.
Specifically, our work focuses and provides in-depth discussion of DNN-specific approximate techniques that are implemented in the hardware level (e.g., logic approximation) and/or modify architecture of the accelerator.}

\begin{table}[t]
\caption{Recent Relevant Surveys}
\label{table:surveys}

\renewcommand{\arraystretch}{1.1}
\footnotesize
\begin{tabular}{
>{\centering\arraybackslash} m{0.15\columnwidth} |
>{\centering\arraybackslash} m{0.1\columnwidth} |
m{0.5\columnwidth}
}
\hline
\textbf{Ref.} & \textbf{Year} & \multicolumn{1}{c}{\textbf{Description/Focus}} \\ \hline
\cite{ApproxCircuitsSurvey} & 2020 & Approximate Arithmetic Circuits. \\ \hline
\cite{Reda:AxBook2018} & 2018 & Approximate Circuits with limited discussion on DNN accelerators with emphasis on software DNN approximation. \\ \hline
%\cite{swagath2020}/\cite{Chen:DATE2018} & 2020/ 2018 & Approximate DNN accelerators with main focus on software DNN approximation. \\ \hline
\cite{swagath2020} & 2020 & \multirow{2}{0.5\columnwidth}{Software-based approximation for DNN accelerators.} \\ \cline{1-2}
\cite{Chen:DATE2018} & 2018 & \\ \hline
\cite{Chen:JENG2020} & 2020 & DNN accelerator architectures \\ \hline
\blue{\cite{Quant_tutorial}} & \blue{2017} &  \multirow{2}{0.5\columnwidth}{Software and hardware optimization for DNN accelerators} \\ \cline{1-2}
\cite{Capra:ACCESS2020},\blue{\cite{Quant_tutorial2}} & 2020 &  \\ \hline
%\blue{\cite{Quant_tutorial2}} & \blue{2020} & \\ \hline
\cite{gholami2021QuantSurvey} & 2021 & Quantization techniques \\ \hline
\cite{liang2021:PrunandQuantSurvey} & 2021 & Software-based Pruning and Quantization techniques \\ \hline
\cite{ren2021ArchitectSurvey} & 2021 & NN architectures \\ \hline
\end{tabular}
\end{table}

\section{Brief Background on Deep Neural Networks}
\label{sec:background}

\begin{figure}[t]
 \centering
\includegraphics[width=1\textwidth]{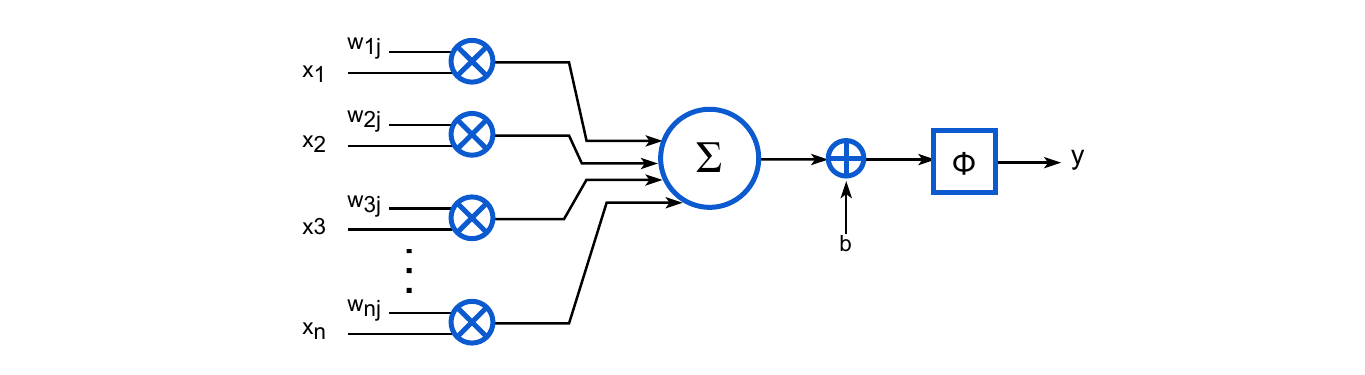}%
\centering
\vspace{-0.5cm}
\caption{
Schematic of a Neuron
}
\label{fig:Neuron}%
\end{figure}

Deep neural networks consist of artificial neurons.
The computation model of a neuron is illustrated in Fig.~\ref{fig:Neuron} and given by~\eqref{eq:neuron}.
Each neuron performs a weighted sum of all its inputs and then a bias term is added for a possible offset~\cite{Capra:ACCESS2020}. 
The result is passed through the activation function, from which the output of the neuron is obtained. 
% The result is passed through a non-linear function $\Phi()$, called activation function, from which the output of the neuron is obtained. 
Neurons are represented as nodes in a graph and are organized in layers.
\blue{In DL, a layer is a function that receives inputs from the previous layers and passes outputs to the next layers~\cite{Goodfellow2016DLbook}.}\label{commentR1C6} 
It is usually uniform, and it only comprises one type of activation function, pooling, convolution etc. 
\begin{equation}
    y_j = \Phi(\sum_{k=0}^{n-1}
    {x_k w_{kj}}
    +b),
    \label{eq:neuron}
\end{equation}
\blue{where $y_j$ is the output of the neuron, $w_{kj}$ are the neuron's weights, $n$ is the number of weights, $x_k$ are the neuron's inputs, $b$ is the bias of the neuron, and $\Phi$ is the activation function.}\label{commentR1C8}

\blue{
The most popular and widely used neural networks today are: Multi-Layer Perceptrons, Convolutional Neural Networks, Recurrent Neural Networks, and Transformers~\cite{google:tpu,swagath2020}.\label{commentR1C1b}
Specifically:
\begin{enumerate}
    \item  \emph{Multi-Layer Perceptrons} (MLPs): Each node in a layer is composed of a nonlinear function of a weighted sum of all the previous outputs (fully connected)~\cite{Quant_tutorial}.
    \item \emph{Convolutional Neural Networks} (CNNs): They are mainly composed of convolutional, pooling, and fully-connected layers and exploit the concept of shared weights and are designed to learn spatial hierarchies of features~\cite{Quant_tutorial}.
    \item \emph{Recurrent Neural Networks} (RNNs): Each layer is composed of nonlinear functions of the weighted sums of the outputs and the previous state. Long Short-Term Memory (LSTM) is the most common RNN. The weights are reused across time steps. A key feature of LSTMs is to decide what to forget and what to forward to the next layer~\cite{lstm}.
    \item \emph{Transformers}: They handle sequential input data as RNNs, but they differ since they use a different mechanism called ``self-attention'' that weights the significance of each input part and enables parallel data processing~\cite{transformer:2019}.
\end{enumerate}
}
\blue{The goal of our work is to survey the state of the art of hardware approximation techniques for DNN accelerators, without any constraints on the DNN type.\label{commentR1C1c}
Though, as it will be shown in Section~\ref{subsec:performanceanal}, the majority of the examined works mainly use only CNNs in their analysis/evaluation.}

% \blue{
% The fundamental operations required for a neural network execution regarding arithmetic computations are mainly multiply-and-accumulate (MACs).\label{commentR3C3} 
% A MAC operation typically multiplies each input with the corresponding weight and adds the previous partial sum to the result. 
% Each of them needs three reads from the memory and one write of data access. Therefore, energy consumption associates with both computation energy, which depends on the number of MAC operations, and data movement.
% Looking at the energy conception, it has been demonstrated that in the worst case where three movements through off-chip DRAM are needed, energy consumption can reach orders of magnitude higher than the required energy for the MAC computation.~\cite{google:tpu}.
% However, also computational requirements in DNNs can be demanding in many AI applications (e.g. smart window), where processing power is continuously needed.
% In this case, common batteries can power such applications only for a few hours~\cite{swagath2020}.
% Thus, new approximate computing techniques are essential for hardware execution of deep networks in a manner that reduces effectively energy consumption and enables on-device implementations.
% }

\subsection{Layers}\label{subsec:layers}

\subsubsection{Fully Connected (FC) Layers}\label{commentR1C2}
In a fully connected layer, the input and output neurons are connected to each other by flattening the matrix into a vector.
Every output neuron performs a weighted sum of every input neuron.
Typically, as convolution layers, FC layers are followed by a non-linear activation and/or bias addition.
FC layers are usually used as the classifier in the final stage of a DNN.
Contrary to convolutional layers, which are compute intensive, FC layers are memory intensive due to the many neuron synapses.

\subsubsection{Convolutional Layers}\label{subsubsec:conv}
This layer carries the main portion of network's computational load.
It performs a dot product between two matrices, where the one matrix is an input feature map and the other is a set of weights known as kernel.
% \blue{In Fig.~\ref{fig:conv}, a simple 2D convolution operation with a $X_k\times Y_k$ \emph{kernel} applied to a single channel input (\emph{input feature map}) to give a $X_o\times Y_o$ output (\emph{output feature map}) is depicted.}
\blue{Fig.~\ref{fig:conv} illustrates the convolution operation between an input of size $[I_x \times I_y \times M]$ and $Z$ filters of size $[K_x \times K_y \times M]$.
The depth of the output (output feature map) is $Z$.}
Once the output feature map is computed, typically the operation of Pooling is performed.
The size of the kernel depends on the size of the receptive field and consequently of the weight matrix.
The distance between adjacent receptive fields is determined by the stride parameter.
All neurons of a layer share the same weight matrix, trying to detect the same feature in different locations of the layer.
To detect multiple features, a convolutional layer has many channels, i.e., many feature maps.
Due to their high computational intensity\footnote{GEMM operations consume more than the 70\% of the inference time of modern DNNs~\cite{swagath2020}.}, convolution layers consist the main approximation target as Section~\ref{sec:techniques} reveals.

\subsubsection{Pooling Layers}
\begin{figure}[t]
 \centering
\begin{subfigure}[t]{0.5\textwidth}
\centering
\includegraphics[width=\textwidth]{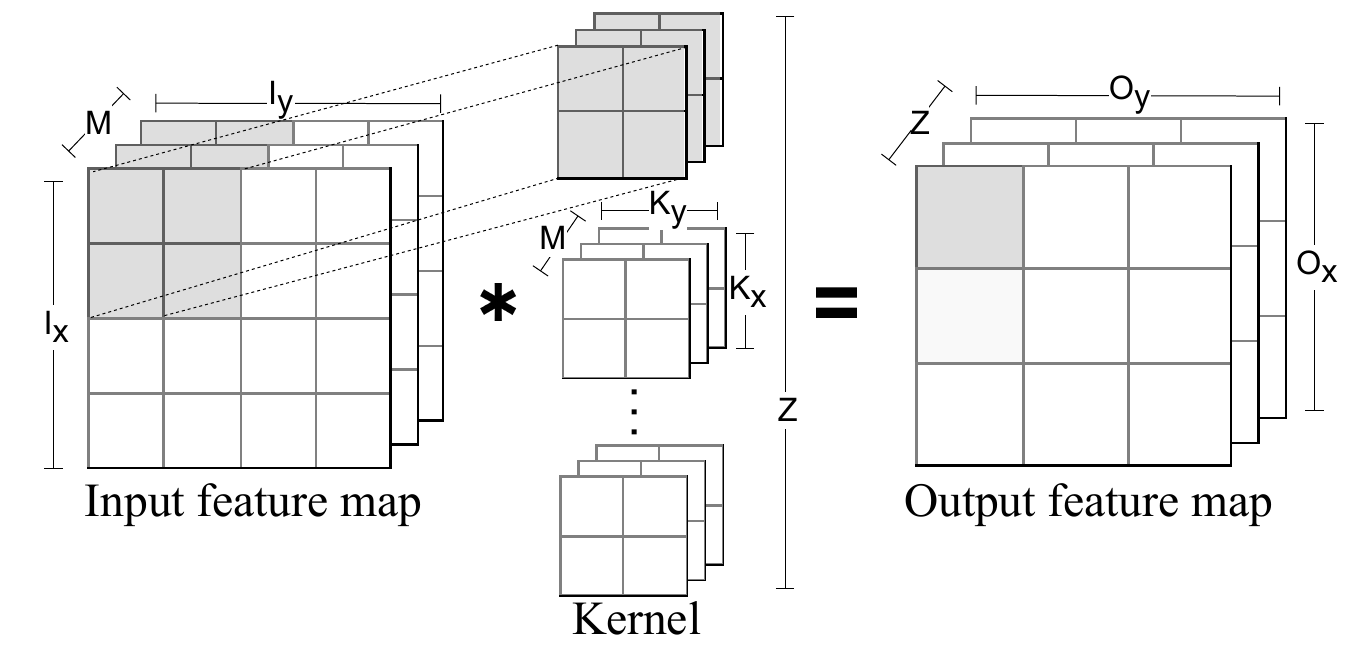}
\caption{\empty}
\label{fig:conv}
\end{subfigure}%
\hfill
\begin{subfigure}[t]{0.5\textwidth}
\centering
\includegraphics[width=\textwidth]{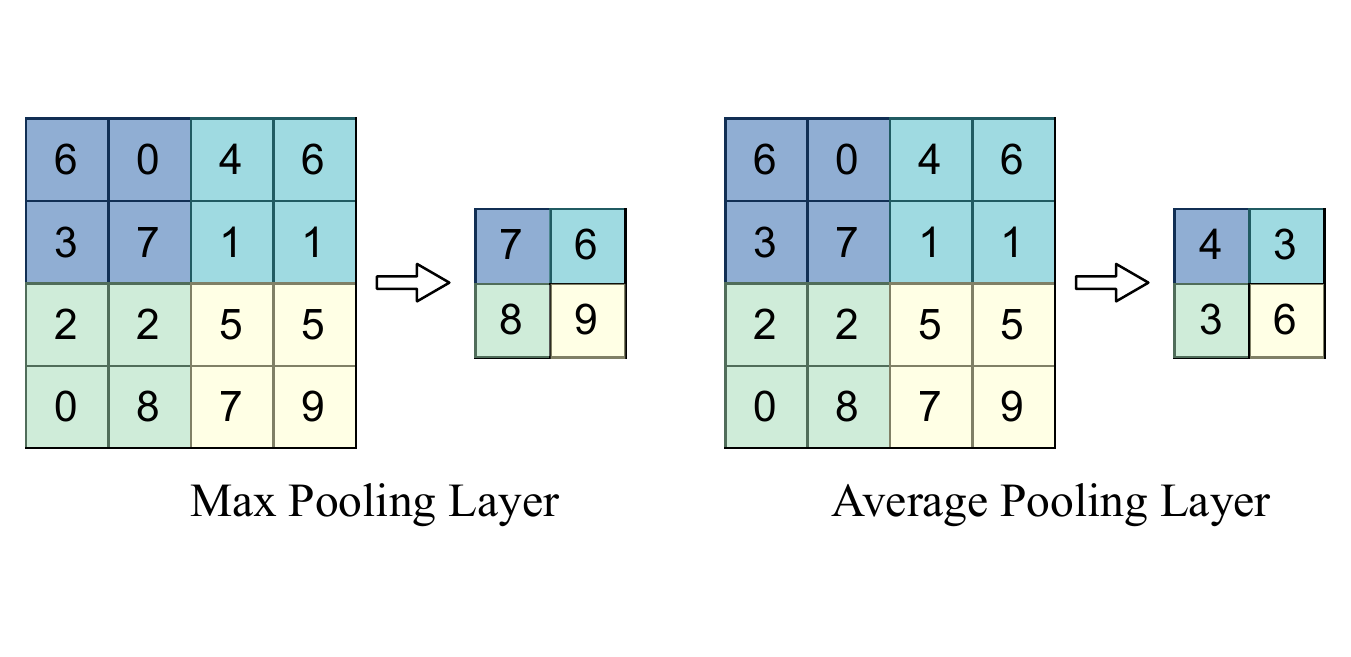}
\caption{\empty}
\label{fig:poolings}
\end{subfigure}%
\caption{ a) Convolution and b) Max Pooling and Average Pooling examples}
\end{figure}

Pooling Layers are placed after the convolutional layers.
Their primary use is to reduce the number of activations of a layer and thus reduce the memory demands and computations needed in the later layers.
This layer substantially down-samples outputs by returning a single value of each group depending on the pooling strategy, e.g., max-/average- pooling (Fig.~\ref{fig:poolings}). 
In max-pooling, the maximum value of the nearby neurons is the output, while in average-pooling the output is their average value.
As Fig.~\ref{fig:poolings} shows, the inputs of the next layers are significantly reduced.
The pooling layer type can be exploited to apply customized approximation (see Section~\ref{sec:techniques}).
% For example, if max pooling is used, then the overall accuracy is defined by the accuracy of the maximum value.

\subsubsection{Activation Functions}
Activation Functions are non-linear transformations that are applied after the weighted sum of the inputs of a neuron.
The activation function increases the fitting ability of NNs and helps solving complex problems that cannot solved with linear algebra. 
The most commonly used activation function is Rectified Linear Unit (ReLU) which forces negative values to be zero and keeps positive values unchanged:
\begin{equation}
{
    y = \left\{
    \begin{array}{l}
         0 \; \; \; \; \;\text{if $x<0$}  \\
         x \; \; \; \; \;\text{otherwise}
    \end{array}
    \right.
    }
    \label{eq:relu}
\end{equation}
Some other activation functions are Sigmoid and TanH, which normalize the output in the range of $(0,1)$ and $(-1,1)$ respectively, while Softmax function normalizes numbers in the range of $(0,1)$ with the restriction that their sum should definitely be equal to $1$.
Many works leverage the activation function to apply optimized approximation  (see Section~\ref{sec:techniques}).
For example, when ReLu is used, the overall accuracy is mainly defined by the accuracy of the positive values\footnote{Without loss of generality accurate sign calculation is assumed.}.

\subsubsection{Normalization Layers}
These layers exploit the fact that neural networks have usually a normal distribution and help keeping input values in the same range. 
The latter speed up the training process and use higher learning rates so that layers do not have to adapt to a different distribution at each training step, making thus learning easier.
A widely used normalization method is Batch Normalization~\cite{BatchNorm}, which transforms $x$ according to the following expression:

\begin{equation}
    y(x) = \gamma \odot \frac{x - \mu_x}{\sigma_x} + \beta,
\end{equation}
where $\mu_x$ and $\sigma_x$ are the mean and standard deviation of the input tensor $x$ and $\gamma$, $\beta$ are respectively the scale and shift parameters.
Those are learned with the rest model parameters during training.

\subsection{Training \& Inference}\label{subsec:traininference}
\blue{\subsubsection{Training}\label{subsec:training}
During training, the network tries to learn the weight values.
A labeled dataset is used for the training process.
A variant of stochastic gradient descent algorithm, which is iterative, is mainly used in training.
The main processes of training are the forward and backward propagation and the weight gradient and update.
In the forward pass, the neurons in each layer are evaluated by traversing all layers in succession from first to last.
In backpropagation, the outputs of the network are compared with the golden outputs and the resulting error is propagated back through the network layers.
The weight update is then performed by accumulating the product of the forward pass activations and the backpropagation errors corresponding to a given weight.
Training is usually executed on distributed systems with many workers and can become a very time consuming procedure.
For example, Facebook required one hour for the $90$ epoch ImageNet training with ResNet-50 using \textit{32 CPUs and 256 NVIDIA P100 GPUs}\yellow{~\cite{facebook:imagenet}.}}\R{comment_minor_R2C2}
%\yellow{Therefore, alternative methods are also explored such as fine-tuning and other statistical approaches (details in Section~\ref{subsec:error}).\R{comment_minor_R3C3}}

A common problem in the training process is overfitting.
Overfitting has not yet \blue{been proven} mathematically but only experimentally and refers to a network that much trained that it produces overly complex and unrealistic class boundaries when data meticulously fits into the model and is memorized.
This leads to poor performance when a new input was never seen before.
Some techniques that help to avoid overfitting by making the model simpler are dropout~\cite{Dropout}, early stopping~\cite{Shao2011:earlystopping}, weight decay~\cite{Leung2010:weightdecay} and learning with noise~\cite{Nagabushan2017:learnwithnoise1}.
As discussed in Section~\ref{sec:techniques}, the approximation noise induced by the approximate circuits might help in mitigating overfitting.

\R{comment_minor_R3C4}\yellow{The works that we studied in our survey have widely employed approximation-aware (re)training (details in Section~\ref{subsec:retraining}) as an error compensation mechanism to mitigate the accuracy loss due to the introduced hardware approximation.
However, given the increased time complexity of training (as mentioned above), retraining can be very time consuming and in the case that approximate hardware emulation is required, the time required can become unsustainable~\cite{Mrazek2019:ALWANN}.
Moreover, it is highly possible that approximation-aware (re)training can be even infeasible, due to proprietary models and/or datasets~\cite{Mrazek2019:ALWANN}.
Therefore, alternative methods are also explored such as fine-tuning and other statistical approaches (details in Section~\ref{subsec:error}).
Finally, it should be mentioned that quantization-aware training has gained a lot of popularity since it enables remarkable model compression and very low-bitwidth integer-only arithmetic inference~\cite{Choi:arxiv2018:pact}.}

\subsubsection{Inference}\label{subsec:inference}
During inference, the already trained NN is used to derive predictions against new unseen data.\label{commentR3C9}
Inference involves only the forward pass.
Training identifies the model parameters while inference uses the model to make predictions.
In contrast to training procedure, inference is typically executed on a single device~\cite{swagath2020} (cloud or even
on a mobile/edge/IoT device) where latency requirements~\cite{google:tpu} as well as energy constraints can become very tight.
Though, the larger a DNN, the more compute and energy is consumed to run inference, and the higher the latency will be.
Hence, although the trained model could be directly deployed to run inference this is rarely the case and several optimizations are examined to meet real world requirements.
To that end, hardware approximation techniques, \yellow{that} constitute the focus of our survey, have been widely studied to enable efficient DNN inference.

\begin{figure}[t]
 \centering
\includegraphics[width=1\textwidth]{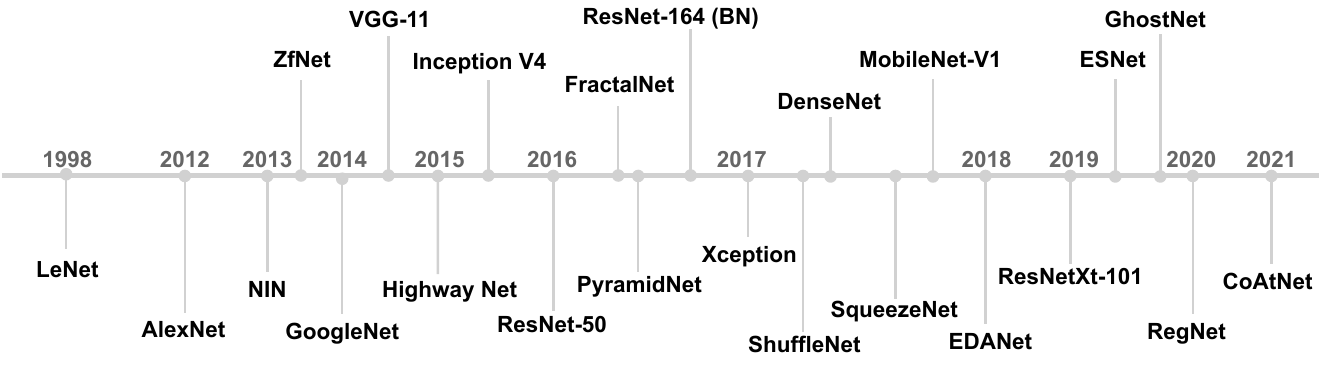}%
\centering
\caption{
Timeline of notable DNNs}
\label{Timeline}
\end{figure}

\subsection{Models and Datasets}

Over the decades, significant research effort has been carried out to improve the performance of DNNs, and particularly CNNs, through novel architectures. 
Fig.~\ref{Timeline} presents some notable CNN models published over time.
CNNs have been applied to vision tasks since 1980s when~\cite{Lecun:1989} proposed a first multilayer CNN named ConvNet. 
LeNet (1998)~\cite{MNIST}, an improved version of ConvNet, achieved significant milestones in recognition tasks.
However, the never ending requirement for higher accuracy led to many new, deeper, and vastly more complex models.

An important aspect in DNNs is the complexity of the task that they have to address.
Datasets are fundamental to test a DNN's accuracy.
Table~\ref{tab:datasets} presents the characteristics of the most commonly used datasets in the works we reviewed in Section~\ref{sec:techniques}.
Many datasets might exist for the same task but different datasets are hardly comparable and their difficulty can significantly vary.
Different datasets reflect to different models and more complex datasets require more complex networks.
The latter translates to more weights and consequently a larger number of operations (MACs).
%Top-1 and Top-5 accuracy are used to test and evaluate a model's performance on a specific dataset.

\begin{table}[t]
\caption{Common datasets used in DNN evaluation}
\renewcommand{\arraystretch}{1.1}
\centering
\footnotesize
\begin{tabular}{c|c|c|c|c|c}
\hline
\textbf{Dataset} & \textbf{Images} & \textbf{Classes} & \textbf{Size} & \textbf{Input Size} & \textbf{Year} \\ \hline
MNIST \cite{MNIST}            & 60K             & 10               & 50 MB        & 28x28  & 1998          \\ \hline
SVHN \cite{SVHN}             & 600K            & 10               & 2.5 GB       & 32x32  &  2011          \\ \hline
CIFAR \cite{CIFAR}            & 60K             & 10/100           & 170 MB       & 32x32  & 2009          \\ \hline
ImageNet \cite{Imagenet}         & 1.5M            & 1000             & 150 GB       & 256x256  & 2009          \\ \hline
\end{tabular}
\label{tab:datasets}
\end{table}

\section{Hardware Approximations for DNNs}\label{sec:techniques}

\begin{figure}[t]
 \centering
\includegraphics[width=1\linewidth]{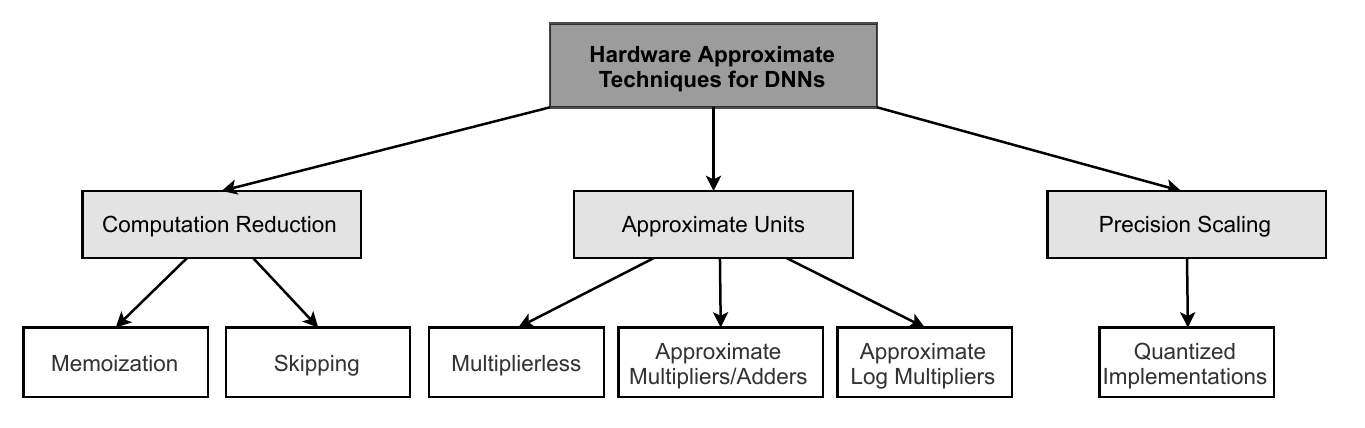}%
\centering
\caption{
Clustering of hardware DNN approximation techniques
}
\label{fig:Clustering}%
\end{figure}

In this section, the state of the art of hardware approximate computing techniques mainly for \blue{deep CNN} inference is discussed.
\blue{Note that although some of these techniques rely on (re)training to mitigate the accuracy loss due to approximation, training is used only as a mechanism to improve the accuracy of the approximate inference and it is not the target of the approximation itself.}\label{commentR3C5}
In addition, after identifying common patterns in examined techniques, we organize them in groups with respect to the type of applied approximation.
%These categories can be used to determine the main approximation methodologies that are applied in DNN accelerators.
As illustrated in Fig.~\ref{fig:Clustering}, hardware DNN approximation can be clustered in three wide categories: \textit{Computation Reduction}, \textit{Approximate (Arithmetic) Units}, and \textit{Precision Scaling}.
It is noteworthy that although these approximation categories are orthogonal, the state of the art applies, mainly, approximations from one category or combines Precision Scaling with approximations from another category.

\begin{figure}[t]
 \centering
\includegraphics[width=1\textwidth]{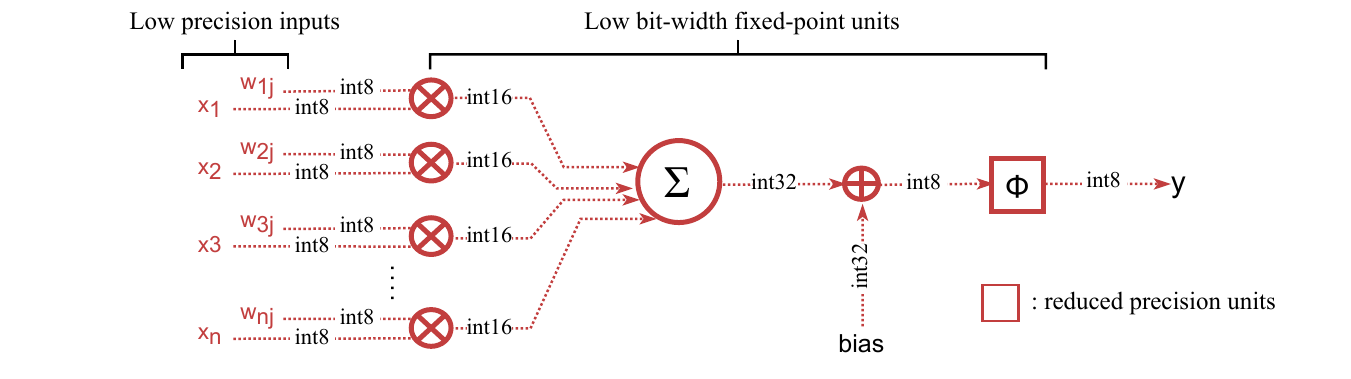}%
\centering
\caption{
\blue{Schematic of a Neuron when applying Precision Scaling approximation.
The precision of all the Neuron’s components is affected (reduced).
This example is adapted from~\cite{Jacob2018:Quant} and illustrates an integer-arithmetic-only $8$-bit inference.}
}
\label{fig:neuron_Quant}%
\end{figure}

\subsection{Precision Scaling}\label{subsec:precision}
Low-precision computation is the key to enable high compute densities in DNN hardware accelerators across cloud and edge platforms.
One of the first and most widely used approximation techniques to enable effective precision scaling is quantization.
% \blue{In integer-arithmetic-only inference weights and activations are quantized to low bitwidth (e.g., $8$-bit) integers and biases are quantized to $32$-bit or lower~\cite{Jacob2018:Quant}.\label{commentR3C6}
% Other quantization approaches mainly target model compression and quantize only the weights, e.g.,~\cite{Zhou2017:INQ}.}
% \blue{
% One of the first and most widely used approximation techniques to enable effective precision scaling is quantization.
% Quantization is being widely used in CNNs to shrink the network size, targeting for high memory savings and simplified computing operations.
% Among our collected works, quantization is preferably performed in convolution layers, rather than fully-connected layers, since the last are typically used in the final stage of the network and the introduced error cannot be balanced, leading hence to higher accuracy loss.  
% Regarding lower block levels than layers, a wide range of data objects can be quantized in a DNN.
% The most commonly quantized objects are weights and activations, which can usually provide significant gains in inference.
% Although bias quantization is also feasible, it is not considered as a common procedure due to its small participation in neural network operations, while accumulation of partial sums is normally done in full precision~\cite{Quant_tutorial2}.\label{commentR1C3}
% Finally, other works study quantization during training, where gradients and errors among all layers are converted to low bit-width values~\cite{wu2018training}.
% }\label{commentR3C6}
Quantized hardware implementations feature reduced bitwidth dataflow and arithmetic units (as illustrated in Fig.~\ref{fig:neuron_Quant}) attaining, thus, very high energy, latency, and bandwidth gains compared to 32-bit floating-point (FP32) implementations.
Traditionally FP32 was used in DNN inference.
Rather than executing all the required mathematical operations with ordinary $32$-bit/$16$-bit floating point (as in CPUs and GPUs), quantization allows us to exploit smaller integer operations instead.
Moreover, quantized implementations reduce the size of the model linearly, leading to high storage gains and low memory transfers.
\blue{In integer-arithmetic-only inference weights and activations are quantized to low bitwidth (e.g., $8$-bit) integers and biases are quantized to $32$-bit or lower~\cite{Jacob2018:Quant}.\label{commentR3C6}
Other quantization approaches mainly target model compression and quantize only the weights, e.g.,~\cite{Zhou2017:INQ}.}
Advancements in quantization methods have demonstrated that integer 8-bit (INT8) DNN inference can achieve almost identical accuracy with FP32~\cite{google:tpu}.
Finally, a significant advantage of quantization is that although it directly impacts the hardware requirements, the accuracy loss is fully controlled and defined at software level.
In other words, the hardware gains will depend only on the supported precision(s) of the accelerator, while the accuracy will depend on the employed quantization method. 
\blue{Though, the latter assumes that the accumulators of the DNN accelerator have enough precision to avoid any overflow and accurately accumulate the partial sums~\cite{Gysel2018:Ristretto,google:tpu,Chen:ISCA2016:eyeriss,Quant_tutorial}.\label{commentR1C3}
If this is not the case, then approximate results may be obtained since the intermediate partial sum might be clipped by a maximum value defined by the precision of the accumulator.
However, the works that we studied in this survey do not consider such an approximation and the size of the accumulator is selected large enough to avoid any overflow, e.g., based on the largest filter size.}
Concluding, studying quantization methods and quantized hardware implementations is out-of-the-scope of this work and comprehensive discussions can be found in many works~\blue{\cite{Chen:DATE2018,swagath2020,gholami2021QuantSurvey,Reda:AxBook2018,liang2021:PrunandQuantSurvey,Choi:arxiv2018:pact,Choi:SySML2019:sawb,Quant_tutorial,Quant_tutorial2}}.
A brief discussion is included in this section for completeness reasons and since many of the approximate techniques discussed hereafter are compatible and/or orthogonal with quantized implementations.
Nevertheless, quantized implementations will not be further analyzed.

At the software side, there are multiple quantization methods and several ways to map the data on the compressed precision levels.
For example:
\begin{itemize}
\item The simplest method is mapping through static fixed-point (fxp) quantization.
A N-bit fixed-point number is represented by $(-1)^s \times m \times 2^{-f} $, where $s$ is the sign bit, $m$ is the (N-1)-bit mantissa, and $f$ is a scale factor.
The energy and area of an fxp multiplier scales approximately quadratically with the number of bits.
For example, an $8$-bit fxp multiply consumes $15.5$x less energy with $12.4$x less area than a $32$-bit fixed point multiply, and $18.5$x less energy with $27.5$x less area than a $32$-bit fp multiply~\cite{energy_comput}.
\item In dynamic fixed-point format, numbers in similar dynamic range are grouped together and share a common fraction length.
This fraction length is chosen based upon the dynamic range of each of the three layers, i.e., inputs, weights and outputs. 
The proper use of this method combined with the examined network's weight analysis can achieve even higher accuracy results than simple fxp method~\cite{GoingDeeper}.
\item Another approach is to use a simple mapping function such as a power-of-two function, where the distance between different quantization levels varies and implementations can be done with simple logic such as a shift operation~\cite{lin2016:PowerOfTwo}.
\item Regarding reduced-precision floating-point numbers, several formats have been explored since the earliest of 2015 \cite{limited_precision}.
Some of them include IEEE FP16 1-5-10, BFloat 1-8-7~\cite{Wang2018:8float} and DLFloat16 1-6-9~\cite{DLFloat} representations.
Hybrid-FP8 (HFP8) supports two formats FP8 1-4-3 and FP8 1-5-2~\cite{swagath2020}, while
Minifloat supports any exponent and mantissa combination~\cite{Gysel2018:Ristretto}.
\end{itemize}

\R{comment_minor_R3C1}\yellow{Hereafter, we present some state-of-the-art quantization techniques that enable the exploitation of precision scaled hardware by mitigating the accuracy loss due to the low numerical precision.
}

\textit{\yellow{Post-training Quantization:}}\label{commentR3C7}
Uniform symmetric~\cite{Krishnamoorthi2018:QuantWhitePap}, asymmetric min/max~\cite{Jacob2018:Quant} are post-training quantization \yellow{(PTQ)} methods and achieve very high accuracy at $8$ bits.
Similarly, a post-training quantization method, ACIQ, is proposed in~\cite{Banner2019:ACIQ}.
ACIQ uses an optimal clipping for quantization which limits the range of activation values in order to reduce the rounding errors while also containing most of the un-quantized information.
In both activations and weights, a bit allocation is applied for each channel to minimize the mean-square-error (MSE).
A bias correction scheme is also introduced to fix the deviation occurred by quantization.
These three methods can be combined to restore most of the accuracy loss without the need of retraining.
% LAPQ~\cite{Nahshan2019:LAPQ} uses different quantization for each layer, in which the quantization step size is optimized to fit in the dynamic range of the values.
% Observations showed semantic interactions between these layers and in order to optimize the loss function of the network with respect to the quantization step size for each layer, an optimization algorithm is also used.
\blue{\cite{pow2quant:date19} presents a \yellow{PTQ} procedure that supports linear quantization for activations and linear, power-of-two, and two-hot quantization for the weights.
In~\cite{pow2quant:date19} the error generated in each layer is used to adjust the quantization step size for both features and weights in an iterative way.}\label{commentR2C4}
Ristretto~\cite{Gysel2018:Ristretto} is an approximation framework that includes dynamic fxp, minifloat, and power-of-two number formats and performs automatic network quantization by evaluating different bit-widths and number representations to find the right balance between compression rate and network accuracy.
% \red{\cite{Zhou2017:INQ}: retraining. \cite{Zhu2017:ternaryquant} fine-tuning requires retraining no? \cite{Choi:SySML2019:sawb} I don't know if sawb needs retrain but it is combined with pact that does.}

% \cite{Choi:SySML2019:sawb} achieves very high accuracy for extremely low bitwidths (less or equal to $2$ bits for weights and activations).
% while for $4$-bit quantized CNNs PACT delivers accuracies similar to FP32 representation.
% Indicatevely, most commonly used works analyzed in this section are presented in Table~\ref{table:LowBitWidth} reporting their validation accuracy compared to 32-bit floating point implementations for different models, datasets and bit-width in weights and activations.
%Of course, a lot of research has been also done in Binarized Neural Networks (BNNs) in which arithmetic operations are replaced with bit-wise operations~\cite{BNN1:2017,BNN2:2016,BNN3:Intel}.
%However, BNNs are a particular category of DNNs, differing from conventional CNNs and they are out of this article’s scope.

\yellow{\textit{Quantization-aware Training:}}\label{QAT}
\R{comment_minor_R3C2}\yellow{Quantization-aware training (QAT) is an approach for training quantized networks.
In QAT, the forward pass simulates the quantized inference while backpropagation is performed as usual and weights and biases are in floating point~\cite{Jacob2018:Quant}.}
% \yellow{QAT} is introduced in~\cite{AxNN:Roy2014} to reflect the quantization errors during the DNN training procedure so that the stochastic gradient descent algorithm can adapt the model towards compensating the errors, \yellow{calculated from backpropagation analysis.}
\yellow{The latter is crucial since accumulating the gradients in quantized precision can result in zero or high error gradients~\cite{gholami2021QuantSurvey}.}
In~\cite{Qil:2019} a quantization method was proposed in which the boundaries of quantization values are parameterized and trained.
Afterwards, values that are smaller than the lower bound are pruned.
The quantizer \yellow{attempts to optimize the trainable parameters with respect to the task loss of the entire network and }can be applied in both activations and weights with extremely low bit-width (2/3/4-bit), achieving state-of-the-art classification accuracies.
\yellow{Additional example is~\cite{LQnets:2018}, where training parameters of the batch normalization layers at high precision is included.}
In~\cite{LQnets:2018} the quantizer \yellow{introduces a perturbation to the model parameters and are jointly trained together, so that model can converge to a point with a better loss~\cite{gholami2021QuantSurvey}.}
The dynamic range and quantization levels can be parameterized in different ways and trained using iterative optimizations.
A weight quantization scheme, statistics-aware weight binning (SAWB), is also proposed in~\cite{Choi:SySML2019:sawb}.
SAWB identifies the optimal scaling factor that minimizes the quantization error based on statistical characteristics of the weights distribution \yellow{(i.e., shape of distribution and representative values throughout the training) }without the need for an exhaustive search.
\cite{Choi:SySML2019:sawb} demonstrates that very high accuracy for extremely low bitwidths (less or equal to $2$ bits for weights and activations) can be achieved.
PArameterized Clipping acTivation (PACT)~\cite{Choi:arxiv2018:pact} uses an activation clipping parameter that is \yellow{learned and optimized via back-propagation }during training to find the right quantization scale.
\blue{PACT demonstrates that although it focuses on activation quantization, also different weight quantization can normally be enabled, delivering accuracies similar to FP32 representation with only 4-bit quantized CNNs.
Consequently, to quantize the weights it uses DoReFa~\cite{dorefa:2018}.
DoReFa is an aggressive and heuristic linear quantization that uses extreme low-precision weights and activations to all layers, excluding only the first and the last ones, \yellow{while gradients are also quantized during the backward pass of the training procedure}.\label{commentR1C10}
}

\yellow{\textit{Binary/Ternary Quantization:}}\label{commentR1C3b}
\blue{
More aggressive precision scaling can be employed to generate \emph{binary} and \emph{ternary} networks.
\R{comment_minor_R3C3}\yellow{Such networks achieve the lowest computational bitwidth and can lead to significant acceleration over higher precisions (e.g., binary arithmetic on NVIDIA V100 GPUs is $8$x higher than INT8~\cite{gholami2021QuantSurvey}).
However, they require customized hardware accelerators to be executed efficiently and training. 
Moreover, for such low precision (binary valued weights), due to the typically small derivatives, it is not effective to update the weights with gradient decent methods~\cite{liang2021:PrunandQuantSurvey}.}
%\yellow{
%Such networks are mapped to different architectures (i.e., bit-wise operations) and can lead to significant acceleration over higher precisions (e.g., binary arithmetic on NVIDIA  V100  GPUs  is 8x higher than INT8~\cite{gholami2021QuantSurvey}.
%}\R{comment_minor_R3C2}
BinaryConnect~\cite{BinConnect:2015} proposed for the first time to use binary weights in \{-1,1\} and~\cite{yodann} used full-precision activations and binary weights.
In Binarized Neural Networks~\cite{bnn} (BNN) and XNOR-Nets~\cite{xnornet} both weights and activations are quantized in binary format.
Such extremely low bit-width formats can replace the costly MAC units by simple XNOR gates followed by pop-count (i.e., count the number of `1').
In~\cite{xnorbin:2018} a CNN accelerator named XNORBIN is proposed with over $25$x higher energy efficiency on competitive models such as AlexNet, while XNOR Neural Engine~\cite{XNORengine} is a configurable hardware accelerator integrated into a microcontroller system, which can fully compute convolutional and dense layers of popular CNNs.
Finally, ternary weight networks use a similar approach but the weights are in \{-1,0,1\}.
}
A ternary quantization is also presented in~\cite{ternary:iclr17}. 
In this approach authors started from a model trained in full precision and then they converted the weights in 2 bits including a fine-tuning process to restore accuracy loss.
% The novelty of this technique is that it consists an asymetric case in which scaling factors of quantization are learned according to current weights and weight gradients.

At the hardware side, low-bitwidth implementations are almost mainstream today.
For example, Eyeriss~\cite{Chen:ISCA2016:eyeriss} and DaDianNao~\cite{Chen:Micro2014:dadiannao} used $16$-bit while Eyeriss V2~\cite{Chen:JETCAS2019:eyerissV2}, Google TPU v1~\cite{google:tpu}, Samsung NPU~\cite{Park:ISSCC2021:samsung6knpu} employ $8$-bit MAC units.
Moreover, many low-bitwidth transprecision architectures are proposed.
Loom \cite{Loom} uses bit-serial multiplicators and both weights and activations have fully variable bit-width, from  $1$~bit to  $16$~bits, while the matrix-matrix multiplication core BISMO~\cite{Bismo} supports precision levels from $8$~bits downto $1$~bit.
BitFusion~\cite{Sharma2018:BitFusion} and BitBlade~\cite{BitBlade} also implement variable precision operations from $1$ up to $16$ bits for DNNs with optimized summations using spatial approach.
%\blue{In~\cite{AdaptDecomp:2021} some fundamental MAC decomposition architectures are presented (vertical and horizontal decomposition) , where further approximation by adapting the maximum possible value of the accumulation with variable error range, is enabled.}\label{commentR2C1}
\blue{In~\cite{AdaptDecomp:2021} fundamental bit decomposition architectures (vertical and horizontal decomposition) are further approximated by constraining the maximum value of the partial sums.}\label{commentR2C1}
IBM RAPID~\cite{Fleischer:VLSIC2018:rapid} uses DLFloat16, $2$-bit (INT2) and $1$-bit fxp while~\cite{Agrawal:ISSCC2021:ibmai7nm} supports DLFloat16 and HFP8 formats as well as INT4 and INT2 formats for highly scaled inference.
Intel Spring Hill~\cite{WechslerHCS2019intel} supports FP16 as well as INT8, INT4, INT2, and even $1$~bit precision operations natively.
Finally, NVIDIA Tensor Cores offer a full range of precisions, i.e., TF32, Bfloat16,  FP16, INT8, and INT4~\cite{Nvidia:TensorCore}.

\subsection{Computation Reduction}\label{subsec:compred}

\begin{figure}[t]
 \centering
\includegraphics[width=1\linewidth]{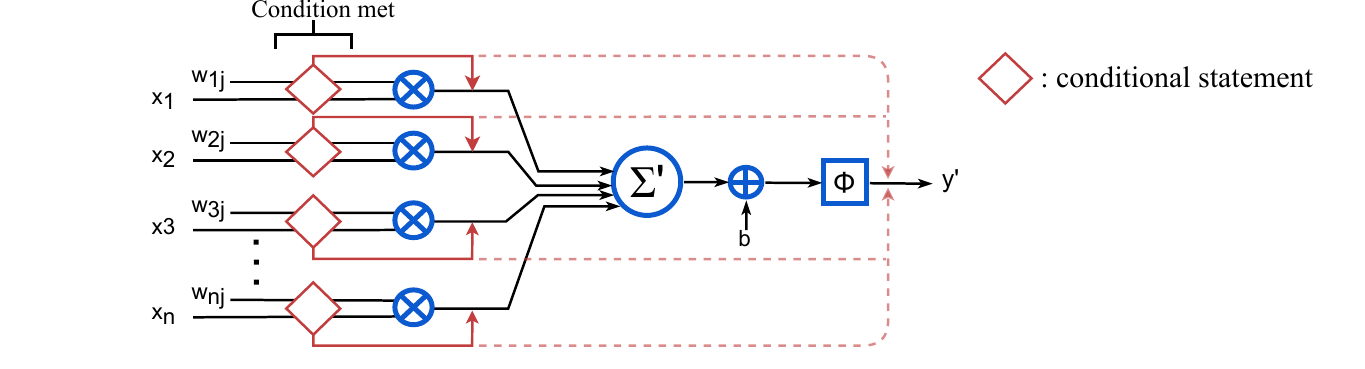}%
\centering
\caption{
\blue{Schematic of a Neuron when applying Computation Reduction approximation. 
A conditional branch is employed to skip or not some of the Neuron's computations.
}}
\label{fig:neuron_CompReduc}%
\end{figure}

During DNN inference, millions of multiplications are performed in the convolution operations~\cite{swagath2020}, leading to high latency and energy consumption, even when considering quantized implementations.
The Computation Reduction approximation category aims in systematically avoiding, at hardware level, the execution of some computations, e.g., multiplications and convolution operations.
As a result, it significantly decreases the executed workload.
% \blue{
% Compared to conventional pruning algorithm, which actually generates a compressed variant of the initial network and is executed offline before inference, computation reduction approximation decides at runtime which operations/computations will not be excecuted.
% }\label{commentR1C4a}
Computation Reduction is further subdivided to the Memoization and Skipping approximation families.
Computation Reduction (illustrated in Fig.~\ref{fig:neuron_CompReduc}) uses a conditional statement to avoid a computation and estimate its output (Memoization) or discard it entirely (Skipping).
\blue{Among the most popular, effective, and extensively used DNN approximation techniques, that reduce the number of the required computations, is the software-based DNN pruning.
Pruning removes connections, filters, and/or channels based on varying importance criteria and can be divided in structured (coarse-grained) and fine-grained pruning.
Pruning actually generates a compressed variant of the initial network and is executed offline before inference.
On the other hand, in the hardware approximation techniques, that we study in this section, the approximation originates from the hardware itself since a conditional statement is integrated in the accelerator and decides at runtime if a computation will be skipped/estimated or not.
Hence, although several architectures exist, e.g., with zero-skipping support, to optimally support the software-based pruning approximation~\cite{EIE:2016, SCNN:2017}, such architectures are not inherently approximate since they will skip computations that do not need to be executed (e.g., multiplication by zero), while the examined Computation Reduction approximation techniques will skip computations that many times should be executed in order to obtain full accuracy.}\label{commentR1C4}

\subsubsection{Skipping}
Skipping approximations aim in reducing the executed workload. 
Such approaches perform a simple computation and evaluate (predict) if a more complex one can be eliminated.
Hence, this approximation family enables \textit{dynamic approximation} at runtime.
The efficiency of the Skipping approximation relies on how often a computation can be skipped, the complexity of the conditional prediction, as well as the complexity of the skipped operation.
%\blue{
%Skipping approximation originates from the hardware itself, since conditional statement is integrated in the accelerator, and differs from zero-skipping architectures~\cite{EIE:2016, SCNN:2017} that maximize the benefits of software-based pruning, but do not affect/approximate the obtained output.
%}\label{commentR1C4b}
Piyasena et al.~\cite{Piyasena2019:PredictandSkip} leverages the widely used ReLu activation function to eliminate redundant computations.
\cite{Piyasena2019:PredictandSkip} estimates the sign of the convolution output using a low-cost prediction scheme.
In this scheme, a power-of-two weight quantization is applied so that multiplications can be replaced with simple logic shifters.
If the estimated sign of the approximate output is negative, the convolution operation is skipped through the clock-gated circuitry, else the original convolution is performed.
\cite{Ujiie2016:SignPredictandSkip} proposes a similar strategy, but the sign estimation is done either after representing weights in ternary format, or after using a sign function, which simplifies the computations, while maintaining the prediction accuracy.
Minkyu Kim et al.~\cite{Kim2021:PSandZS} exploits the max-pooling layers and adopts a precision-cascading scheme to predict and calculate only the maximum value of a convolution operation.
This technique, combined with a zero-skipping scheme, can efficiently avoid redundant computations without affecting neuron synapses that contribute a lot in classification accuracy.
\blue{In~\cite{Hemmat2020a:AirNN}, the weights of each layer of a given CNN are clustered offline in groups. K-means is used for clustering and weights within a cluster feature the highest similarity to each other while weights of different clusters exhibit the least similarity.\label{commentR3C8}
During inference (i.e., at the runtime) only some weight-groups are used while the weights of the rest groups are assumed to be zero.
Finally, the difference of the two output neurons with the highest values is calculated.
If the difference is above a given threshold, the obtained prediction is the output of~\cite{Hemmat2020a:AirNN}, else the inference is repeated using gradually more weight-groups.}
% In~\cite{Hemmat2020a:AirNN}, K-means is used to cluster offline the weights of each layer in groups so that weights within a cluster feature the most similarity to each other, while weights in different clusters exhibit the least similarity.
% Then DNN inference is executed using a fraction of the total weight-groups based on a calculated score.
% If the score is above a given threshold execution terminates else more iterations are performed using additional weight-groups.
% Again, exploiting ReLu, the groups with the maximum positive and minimum negative average weights across the clusters are preferred.
Finally, Huan et al.~\cite{Huan2016:NearZero} introduced Near Zero Approximation (NZA).
NZA exploits the fact that when the multiplication operands are very small (close to zero) the product will be almost zero.
\cite{Huan2016:NearZero} counts the leading zeros of the multiplication operands and if their number is above a threshold, the product is assumed zero and the multiplication is skipped.

\subsubsection{Memoization}
The second subcategory of the Computation Reduction is Memoization.
Memoization avoids a computation (e.g., multiplication or convolution) by replacing its output with the output of a previously performed similar computation.
Hence, the efficiency of this approach depends on the input similarity (i.e., how often a replacement takes place) as well as the complexity of the eliminated computation.Jiao et al.~\cite{Jiao2018:BloomFilter} applies Memoization through a configurable Bloom Filter (BF) unit that stores the product of frequently computed patterns and avoids performing the respective multiplications.
A memoization set of 3000 images was used in~\cite{Jiao2018:BloomFilter} to identify such patterns.
% \red{In~\cite{2019SkippingCNN} authors follow a similar methodology, but instead of BFs they use memory lookups and they eliminate redundant computations repeated multiple times by a range-based clustering algorithm. DELETE?}
Mocerino et al.~\cite{Mocerino2019:CompReuseFP} proposed a CAM-enhanced floating-point unit (FPU) to implement Memoization.
Pre-computed multiplication results are reused whenever a similar input pattern occurs, avoiding thus unnecessary computations of frequent operations.
To increase the frequency of patterns, a clustering approach based on the Jenks Natural Breaks algorithm is applied to weights and activations.
The processing unit of~\cite{Mocerino2019:CompReuseFP} is pipelined and consists of two CAMs (one for the weights and one for the activations) and an SRAM.
If the input pattern is pre-computed, the product is loaded form the memory and the multiplier is avoided by clock-gated signals.
On the other hand,~\cite{InpSimilarity:2018} showed that more than $60\%$ of the inputs of network layer exhibit negligible changes with respect to the previous execution.
Based on that fact, they proposed a method to reuse some results from the previous execution, avoiding all the computations associated with those results.

\subsection{Approximate Units}\label{subsec:axunits}
\begin{figure}[t]
 \centering
\resizebox{.9\textwidth}{!}{%
\includegraphics[width=1\linewidth]{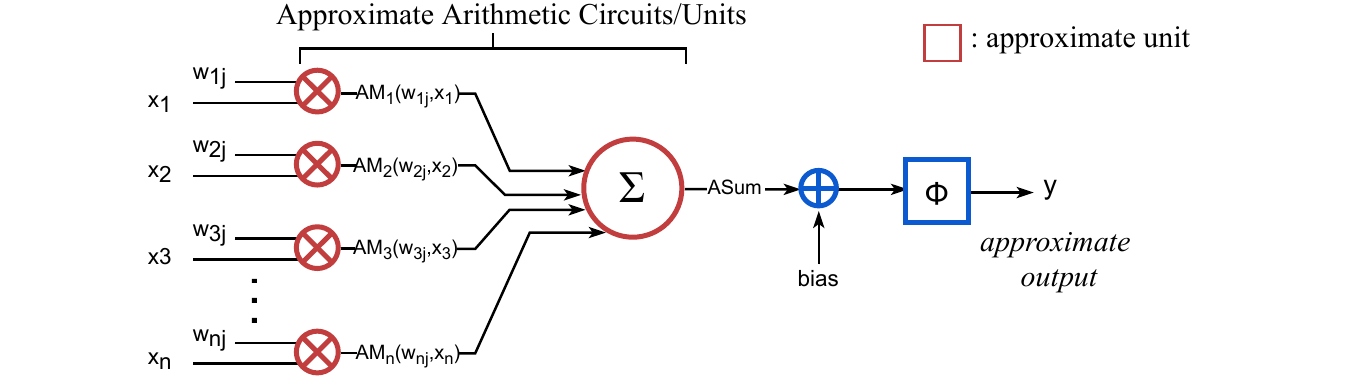}%
}%
\centering
\caption{
\blue{Schematic of a Neuron when applying Approximate Units approximation. 
The multiplication and/or addition units of the Neuron are replaced by approximate ones.}
}
\label{fig:neuron_ApproxUnit}%
\end{figure}

DNN hardware accelerators comprise thousands of multiply-accumulate (MAC) units~\cite{google:tpu}.
This wide category improves the energy consumption and/or latency of DNN accelerators by employing approximate circuits that replace accurate MAC units (Fig.~\ref{fig:neuron_ApproxUnit}).
%As a result, the obtained energy savings are proportional to the number of the replaced MAC units.
%The scope of such circuits is to exploit the tolerance of CNNs to inaccuracies, while also satisfying constraints given by the user.
Approximate Units can be further divided into three approximation families: \textit{Approximate Multipliers/Adders}, \textit{Multiplierless}, and \textit{Approximate Log-Multipliers}.
Briefly, Approximate Multipliers/Adders modify the circuit implementation of the multiplier/adder (e.g., logic approximation), Multiplierless replaces the multiplication \yellow{with a simpler} operation (e.g., addition), and the Approximate Log-Multipliers family replaces the exact binary multiplier with a logarithmic multiplier that is further approximated.

\subsubsection{Approximate Multipliers/Adders}\label{subsec:approxmults}
Considering the vast number of MAC operations required in the inference phase, several works focus on approximating the circuit of the MAC unit itself.
Exploiting a constant energy gain per MAC operation performed, very high energy gains are obtained at inference level.
Targeting approximate MAC circuits, state of the art mainly approximates the multiplier, since it is more complex and power consuming than the adder~\cite{Zervakis:TVLSI2016:perf,Mrazek2017:Evo8,Shafique:DAC2015:gear,approxWallace:henk}.

Mrazek et al.~\cite{Mrazek2016:CGPevo} employ a Cartesian genetic programming (CGP) based optimization -- since it is intrinsically multi-objective and produce efficient approximate arithmetic circuits~\cite{Mrazek2017:Evo8} -- to generate approximate multipliers for inference accelerators~\cite{Mrazek2016:CGPevo}. 
The multipliers generated by~\cite{Mrazek2016:CGPevo} satisfy a given worst-case error constraint and ensure that multiplication by $0$ is always accurate.
An iterative optimization procedure is used to identify the error constraint for the generation of approximate multipliers in CGP-optimization so that an inference accuracy loss threshold is satisfied.
During the iterative procedure, after replacing the accurate multipliers with the approximate ones, the network is retrained to obtain the best quality results~\cite{Mrazek2016:CGPevo}.
Similarly, Vasicek et al.~\cite{Vasicek2019:DataDistrib} use CGP-based optimization to generate approximate multipliers.
In order to avoid time consuming CNN evaluation during the optimization phase, \cite{Vasicek2019:DataDistrib} used the Weighted Mean Error Distance (WMED) metric to quantify the accuracy of the approximate multipliers.
To calculate WMED, the significance of each error is determined by the probability mass function of the network's weight distribution.
Ansari et al.~\cite{Ansari2020:ImprovEvo} evaluated $600$ approximate multipliers ($500$ CGP-based ones
and $100$ variants of deliberately designed multipliers) in CNN inference showing that they can deliver significant gains in terms of power and area for a minimal accuracy loss.
Moreover,~\cite{Ansari2020:ImprovEvo} discussed that the induced approximation noise helps to mitigate the overfitting problem, and thus can even improve the obtained accuracy.
After analyzing $600$ approximate multipliers, a significant conclusion of~\cite{Ansari2020:ImprovEvo} \yellow{showed} that when designing approximate multipliers for CNN inference, the most important error metrics are the error variance and the root mean square error.
Similar to~\cite{Mrazek2016:CGPevo},~\cite{Vasicek2019:DataDistrib} and~\cite{Ansari2020:ImprovEvo} apply retraining to mitigate the accuracy loss due to the approximate multiplications.
\blue{Nevertheless, approximation-aware retraining can be very time consuming as discussed in Section~\ref{subsec:traininference}}.

Mrazek et al.~\cite{Mrazek2020:Libraries} extended the EvoApprox8b library~\cite{Mrazek2017:Evo8} and generated $8\times N$-bit approximate multipliers. 
CGP-optimization and the quality metric of~\cite{Mrazek2016:CGPevo} is used for the generation of the approximate multipliers.
\cite{Mrazek2020:Libraries} evaluated the generated approximate multipliers in CNN inference.
Through a comprehensive analysis,~\cite{Mrazek2020:Libraries} demonstrated that for less complex CNNs (ResNets~\cite{ResNet} on CIFAR10), approximate multipliers may deliver considerable power savings for minimal accuracy loss (even without retraining).
A similar approach that aims to eliminate multiplications by quantizing one term in power-of-two format, is presented in~\cite{lin2016:PowerOfTwo}.
In this method, during the forward pass weights are converted in ternary format, while in back propagation weights and activations are quantized up to 4 bits to improve the accuracy.
Nevertheless, this is not the case for more complex CNNs (ResNet-164 on CIFAR100) where even for $10\%$ energy reduction the accuracy loss is considerable.
Leveraging that weights are known after training, CAxCNN~\cite{Riaz2020:CaxCNN} uses the Canonic Sign Digit (CSD) representation to encode the weights.
CSD uses ternary form $\{-1,0,1\}$ and to represent a binary number, CSD features the least number of non-zeros $\{-1,1\}$.
In addition, adjacent bits cannot be both non-zero.
Exploiting these two features of CSD,~\cite{Riaz2020:CaxCNN} applied truncation and generated approximate CSD multipliers with very small footprint as well as low latency.
Although~\cite{Riaz2020:CaxCNN} performs an optimization search to identify the optimal truncation parameter, CAxCNN does not require retraining.
Exploiting that different layers feature varying resilience to approximation, ALWANN~\cite{Mrazek2019:ALWANN} applied a non-uniform approximation.
ALWANN generates an heterogeneous DaDianNao architecture~\cite{Chen:Micro2014:dadiannao} by using heterogeneous processing elements (PEs).
The employed PEs are built upon different approximate multipliers from the EvoApproxLib~\cite{Mrazek2017:Evo8}.
ALWANN~\cite{Mrazek2019:ALWANN} implements a layer-wise approximation in which each layer is mapped to a specific PE type.
A genetic optimization procedure is used to identify the approximate multiplier per PE as well as the layer mapping to PEs.
ALWANN avoids retraining and recovers some of the accuracy loss by employing a simple, approximation aware weight-tuning procedure.
Similarly, Zervakis et al.~\cite{Zervakis:ACCESS2020} applied also layer wise approximation.
\cite{Zervakis:ACCESS2020} used wire-by-switch replacement to generate an approximate multiplier with three accuracy (relative error) modes.
Hence, using this reconfigurable multipliers, \cite{Zervakis:ACCESS2020} generated an homogeneous approximate architecture.
Through an exhaustive exploration, \cite{Zervakis:ACCESS2020} determined the accuracy mode per convolution layer and generated the respective accuracy-energy consumption Pareto front.
Tasoulas et al.~\cite{Tasoulas2020:weightoriented} introduced the weight-oriented approximation.
\cite{Tasoulas2020:weightoriented} generated a low-variance approximate multiplier (LVRM) with three approximation modes (i.e., three error variance values).
An greedy procedure is used in~\cite{Tasoulas2020:weightoriented} to map weight ranges to the approximation modes of LVRM.
The significance of each convolution layer is also used in the mapping procedure, i.e., weights of less sensitive layers are entirely mapped to the highest approximation.
In addition,~\cite{Tasoulas2020:weightoriented} proposed a bias-correction method in order to avoid retraining and mitigate the accuracy loss due to the approximate multiplications.
Hammad et al.~\cite{Hammad2021:controller} performed approximate multiplication using the Dynamic and Static Segmented Multipliers (DSM, SSM), which perform the multiplication with $m$-bit input segments (where $m$ is smaller than the input bit-width).
In SSM the most significant segment that contains an `1' is used (static) while in DSM the segment is dynamically selected based on a leading one detector (LOD).
To attain high accuracy, \cite{Hammad2021:controller} generated a reconfigurable accelerator that comprises low precision (low $m$) and high precision (high $m$) approximate multipliers.
A low cost classifier is trained to predict the required precision (low or high) for each input image.
At runtime, a controller decides the precision level and then inference is executed using the respective approximate multipliers.
%Guo et al.~\cite{Guo2020:DoubleModeMult} employed low bit-width quantization and proposed a reconfigurable architecture to maximize the hardware utilization under low precision.
Guo et al.~\cite{Guo2020:DoubleModeMult} proposed an approximate multiplier that can support one 16- by 8-bit multiplication or two 16- by 4-bit multiplications and uses an approximate adder to add/merge the outputs of the sub-multiplications.
The proposed approximate adder extends the block-based adder GeAr~\cite{Shafique:DAC2015:gear}.
\cite{Guo2020:DoubleModeMult} observed that in a quantized \blue{CNN}, the inputs of the multipliers roughly follow a Gaussian distribution instead of a uniform distribution.
Exploiting the correlation of the bits for Gaussian distributed inputs, \cite{Guo2020:DoubleModeMult} generated approximate adders with an unequally sized block structure to trade-off between accuracy and circuit delays.
\cite{Guo2020:DoubleModeMult} considers an Eyeriss-like architecture~\cite{Chen:ISCA2016:eyeriss} that uses the proposed approximate reconfigurable multipliers and employs different quantization precision for different layers (i.e., 8-bit or 4-bit).
Exploiting the proposed reconfigurable approximate multiplication, layers with 4-bit weights are executed at higher throughput.

Hanif et al.~\cite{Hanif2019:CANN} considers a systolic MAC array architecture~\cite{google:tpu} and introduces a curable approximation technique.
\blue{``Curable'' approximation refers to approximation approaches that feature an internal error compensations mechanism that enables them to self-correct the induced error.\label{commentR3C10}
This is mainly achieved by estimating the error at runtime and compensating it at a later stage.}
CANN~\cite{Hanif2019:CANN} splits the adder of the MAC unit in two parts (low and high) and cancels the carry propagation from the low to the high part.
Hence, the carry chain (and thus the delay) of the MAC unit is decreased.
To cure the introduced error, the output carry of the low part is accumulated in the next cycle by the neighbouring MAC unit.
The errors generated by the eliminated carries of the border MAC units are not cured.
Zervakis et al.~\cite{Zervakis2021:ControlVar} considered also a systolic MAC array architecture~\cite{google:tpu} and replaced the accurate multipliers with the approximate perforated ones~\cite{Zervakis:TVLSI2016:perf}.
The perforated multipliers omit the generation of some partial products and thus the induced error is known apriori~\cite{Zervakis:TVLSI2016:perf}. 
\cite{Zervakis2021:ControlVar} introduced a control variate approximation technique to heal the approximate multiplication error at runtime.
\cite{Zervakis2021:ControlVar} leverages that the weights are known after training and that the error of the perforated multipliers can be rigorously expressed in order to formulate a control variate that efficiently estimates the runtime convolution error based on the values of the input activations.
An additional column of MAC units is required to accumulate the control variate and compensate the error.

\blue{Concluding, the integration of Approximate Multipliers/Adders in neural network accelerators has attracted significant research interest over the last years.}\label{commentR3C11}
The approximation techniques that belong in this family can be further organized as follows:
\begin{itemize}
\item ~\cite{Mrazek2016:CGPevo,Vasicek2019:DataDistrib,Ansari2020:ImprovEvo,Riaz2020:CaxCNN,Mrazek2019:ALWANN} generate NN-specific approximate accelerators, i.e., apply NN-specific approximations.
On the other hand, \cite{Mrazek2020:Libraries,Zervakis:ACCESS2020,Tasoulas2020:weightoriented,Hammad2021:controller,Guo2020:DoubleModeMult,Hanif2019:CANN,Zervakis2021:ControlVar}  generate generic approximate accelerators.
\item \cite{Mrazek2016:CGPevo,Vasicek2019:DataDistrib,Ansari2020:ImprovEvo} apply retraining to mitigate the accuracy loss, while~\cite{Mrazek2020:Libraries,Riaz2020:CaxCNN,Mrazek2019:ALWANN,Zervakis:ACCESS2020,Tasoulas2020:weightoriented,Hammad2021:controller,Guo2020:DoubleModeMult} 
do not require/apply retraining, and~\cite{Hanif2019:CANN,Zervakis2021:ControlVar} employ a runtime curable approximation technique.
\item~\cite{Mrazek2016:CGPevo,Vasicek2019:DataDistrib,Ansari2020:ImprovEvo,Riaz2020:CaxCNN,Mrazek2020:Libraries,Hanif2019:CANN,Zervakis2021:ControlVar} apply static approximation while~\cite{Mrazek2019:ALWANN,Zervakis:ACCESS2020,Tasoulas2020:weightoriented,Hammad2021:controller,Guo2020:DoubleModeMult} employ dynamically reconfigurable approximate architectures.
\end{itemize}

\subsubsection{Multiplierless}
The Multiplierless subcategory aims in maximizing the gains by eliminating the expensive multiplication circuits.
To achieve this, multipliers are replaced by circuits that implement a simpler operation. 
Parmar et al.~\cite{Parmar2020:CORDIC} exploited the fact that scaling the input feature map does not affect the features extracted by max-pooling and reduced the span of the scaled weights to $[-1, 1]$.
This condition allowed to introduce in the convolution equation trigonometric functions,  which can be implemented by the low-cost CORDIC algorithm.
In~\cite{Faraone2020:AddNet}, authors proposed reconfigurable constant coefficient multipliers (RCCM) that use only adders and shifters.
The supported coefficients are extracted offline based on a distribution matching technique that allows specific RCCM to be selected depending the model's weights.
Sarwar et al.~\cite{Sarwar2018:Alphabet} employed multiplierless neurons by replacing multipliers with simplified shifts and add operations controlled by a unit.
The so called Alphabet Set Multipliers (ASM) comprise a pre-computer bank to compute lower-order multiples of the input based on some small-bit sequences termed alphabets ($\{1, 2, 3, 5, ...\}$), an adder, and one or more select, shift, and control logic units.
The size of the alphabet defines the accuracy as well as the energy benefits of ASM.
An efficient retraining is \yellow{finally} performed in order to tune the weights and mitigate the accuracy degradation due to ASMs.

\subsubsection{Approximate Log-Multipliers}
The Log-Multipliers subcategory converts multiplications into additions by taking approximate logarithm.
Mitchell~\cite{Mitchell1962} proposed an approximate multiplier that employs the log multiplication property.
\cite{Mitchell1962} proposed to compute approximate binary log and antilog by a linear approximation of the log-antilog curves between each power-of-two-interval.
Saadat et al.~\cite{Parames2018:FPMult} extended~\cite{Mitchell1962} to generate a minimally biased approximate multiplier.
\cite{Parames2018:FPMult} observed that in Mitchell's algorithm the error value is always negative.
Through a mathematical analysis,~\cite{Parames2018:FPMult} demonstrated that with the addition of a constant correction term, overall the error is reduced and the average error is pushed close to zero.
In addition,~\cite{Parames2018:FPMult} applied truncation to reduce the size of the main components required (i.e., adder and barrel shifters).
Kim et al.~\cite{Kim2019:mitchell} optimized Mitchell's logarithmic multiplier for approximate CNN inference.
\cite{Kim2019:mitchell} improved Mitchell's implementation (LOD, shift, and decoder blocks), introduced a zero-checking block, that is mandatory to improve the performance of CNNs, and further approximated the design by applying truncation and one's complement for negation.
In~\cite{TwoStageLog:2021}, another approximate logarithmic multiplier with two stages of approximations was proposed. 
During the first stage, the two operands are split into two parts and the proposed multiplier selects either the upper part (if it contains at least one non-zero bit) or the lower part (if not), for the following computations.
The second stage of approximation concerns the binary-to-logarithm conversion, where, in order to reduce more the complexity of circuitry, only a number of (leftmost) bits in the mantissa part are kept.
Vogel et al.~\cite{Vogel2018:ArbitraryLog} introduced a quantization scheme to fractional powers-of-two (e.g., $2^{1/4}$) and showed that the latter provides higher resolution at higher values and the granularity of weight distribution becomes more fine-grained. 
Based on the proposed quantization,~\cite{Vogel2018:ArbitraryLog} replaced the binary MAC units with logarithmic processing elements (PEs) that use an adder, a lookup table (for the required exponents), and a barrel shifter before accumulating the result.

\section{Error Compensation Techniques}\label{sec:CompTechniques}

Although DNNs feature an inherent error resilience, naive or aggressive approximation may result in unacceptable accuracy loss.
In addition, complex networks can become very sensitive to even slight approximation~\cite{Mrazek2020:Libraries,Tasoulas2020:weightoriented,Zervakis:ASPDAC2021}.
In this section, we discuss techniques employed by the state of the art to achieve high accuracy albeit the applied approximations.
Such techniques enable satisfying tight accuracy constraints and/or increasing the applied approximation to further boost the attained gains.

\subsection{Accuracy Recovery with Retraining}\label{subsec:retraining}

Mrazek et al.~\cite{Mrazek2016:CGPevo} showed that when approximate multiplication is used instead of accurate one, the classification accuracy of the examined network decreased to almost $10\%$.
\blue{However, \cite{Mrazek2016:CGPevo} applied approximation-aware retraining and after only $5$ epochs, the accuracy was recovered to more than $90$\% for MNIST dataset.}\label{commentR2C3a}
The backpropagation algorithm was employed in~\cite{Mrazek2016:CGPevo} using the approximate multipliers in the forward pass.
A similar approach is followed in~\cite{Ansari2020:ImprovEvo}.
Despite the high accuracy achieved, retraining can become very time consuming since i) retraining large NNs can be very slow and ii) in the feedforward it requires emulation of the approximate hardware.
To accelerate the accuracy evaluation when using approximate multipliers, \cite{Vaverka2020} proposed TFApprox, a GPU-based hardware emulation framework that extends TensorFlow and supports approximate multiplication through lookup tables.
%The flow diagram of retraining process that is followed by most of collected works is illustrated in Fig.~\ref{fig:retraining}.

A hindering factor to efficient retraining can be the non-uniformity of the approximate multipliers.
Although a proper learning rate, i.e., a multiplication factor in the weight update equation~\cite{Palm2012learningRate}, can efficiently adjust network with approximations in place, many approximate multipliers may require careful regulation.
For example, in the ASM multipliers~\cite{Sarwar2018:Alphabet} using one alphabet, the allowed weight levels are $0\times$, $1\times$, $2\times$, $4\times$ and $8\times$.
The distance between $2\times$ and $4\times$ is $2\times$, while the distance between $4\times$ and $8\times$ is $4\times$.
In this case, a low learning rate would not be enough for weights to be updated properly and overcome the distance barrier between allowed levels.
This would cause weights to condense in a specific level, resulting in a high network accuracy loss.
The same effect would have a high learning rate, too.
Hence,~\cite{Sarwar2018:Alphabet} used initially the highest learning rate that was used to train the \blue{CNN} without approximation.
If the accuracy improves and satisfies given constraints, retraining is carried on with the same learning rate for a few more iterations and until no significant improvement in the accuracy is observed. 
If the accuracy does not improve, the learning rate is reduced by a factor and the approximate \blue{CNN} is further retrained. 
This process of regulating the learning rate is continued until the accuracy improvement saturates.
Beyond the conventional backpropagation algorithm, some works have proposed extended formulations for approximate DNN retraining.
AxTrain~\cite{AxTrain} is a hardware-oriented framework for DNN training which supports approximate inference.
In~\cite{AxTrain} two DNN training techniques were introduced, referred as \textit{passive} and \textit{active} methods.
During retrain a stochastic error model is back propagated to the network parameters in order to minimize the noise sensitivity and the network's accuracy.
Substantially, passive method concerns the training procedure, trying to recover accuracy loss, while active method helps the network to learn the noise distribution with minor modifications in each epoch and become more robust to hardware approximations.
A novel regularization~\cite{Regularization} method, called \textit{alpha regularization}, to bias the training algorithm for approximate \blue{CNN} was presented in ProxSim~\cite{ProxSim}.
Similarly with passive training of AxTrain, ProxSim simulates approximate hardware elements during the computations of \blue{CNN} forward pass and then a regularization term is added to minimize the propagated approximation error for each \blue{CNN} layer.
Although this method was about $2\%$ slower than AxTrain, it appeared to be more efficient in $90\%$ of the cases, as it delivered better improvements in accuracy.
Considering implementations with large approximation errors,~\cite{Distillation:Date21} presented a novel methodology for efficient error recovery through Knowledge Distillation (KD)~\cite{KnowDistill} for approximate DNNs.
This methodology consists of two stages in which firstly FP information are distilled into a quantized model and then into a approximated model.
% In combination with this KD method, they proposed also to estimate the gradient of approximate GEMMs, used for convolution and FC layers, more precisely.
This recovery error scheme achieved a small accuracy loss (<3\%), having energy savings but no improvement in retraining time.
On the contrary,~\cite{Kumar2021:retrain} achieved a reduction in retraining time of up to $11\times$, compared to a behavioral simulation of approximate multipliers in DNNs using ProxSim.
In~\cite{Kumar2021:retrain} an obtained error model was added to each NN layer before activation function, targeting to an improvement in the DNN generalization.
This generalization can be considered as a regularization method, which leads to better and faster results than training with the behavioural simulation.

% Their method includes a model for approximation errors introduced by approximate hardware that identifies the most critical elements affected by the approximation.

% Figure~\ref{fig:retraining} illustrates the flow diagram of retraining process, which begins with the approximate network that is retrained to achieve best quality results.
% The tested accuracy $M$ is checked if it is improved, while meeting also the user's accuracy constraints and if 

% \begin{figure}[ht]
%  \centering
% \includegraphics[width=1\linewidth]{FIGURES/retrain.pdf}%
% \centering
% \caption{
% Flow Diagram of Retraining Process
% }
% \label{fig:retraining}%
% \end{figure}

\subsection{Statistical Error Compensation}\label{subsec:error}

Several works investigate alternatives to retraining in order to improve the accuracy achieved.
As aforementioned, retraining is time consuming and might not always be feasible (e.g., proprietary datasets).
ALWANN~\cite{Mrazek2019:ALWANN} proposed a fast weight-tuning algorithm that adapts the weights according to the employed approximate multiplier and does not require any preprocessing or inference evaluation.
\cite{Mrazek2019:ALWANN} replaced the weights in each layer based on the error characteristics of the employed approximate multiplier.
Each weight $w$ was replaced by $w^\prime$ as follows:
\begin{equation}\label{eq:alwann}
  \argmin_{\forall w'} {\sum_{\forall \alpha}|{M_{ax}(a,w') - a\cdot w}|},
\end{equation}
where $M_{ax}$ corresponds to the approximate multiplication.
Using~\eqref{eq:alwann},~\cite{Mrazek2019:ALWANN} selected the value $w^\prime$ that minimizes the sum of absolute differences (error) between the output of the approximate and accurate multiplication over all inputs ($\forall \alpha$).
In other words, given an approximate multiplier,~\cite{Mrazek2019:ALWANN} updated the weights so that the Mean Error Distance (MED) of the performed approximate multiplications is minimized. 

The main computation of a convolution operation is given by:
\begin{equation}
    Y_o = \sum_{i=1}^{N}{W_{o,i}X_i} + b_o,
    \label{Neuron_Equation}
\end{equation}
where $W_{o,i}$ are the filter's weights, $X_i$ are the input activations, and $N$ is the number of weights.

The error ($\epsilon$) of an approximate multiplier can be viewed as a random variable defined by its mean value $\mathrm{E}[\epsilon]$ and its variance $\mathrm{Var}(\epsilon)$~\cite{Tasoulas2020:weightoriented}.
Therefore, if the approximate multiplication error is systematic, it can be compensated by a constant correction term~\cite{Tasoulas2020:weightoriented}.
Given the convolution operation~\eqref{Neuron_Equation} and following this reasoning,~\cite{Tasoulas2020:weightoriented} proposed a bias-update method to encompass this correction term and compensate, thus, the error induced by the approximate multiplications.
Tasoulas et al.~\cite{Tasoulas2020:weightoriented} proposed to replace the bias term $b_o$ in~\eqref{Neuron_Equation} by $b_o^\prime$.
The latter is given by:
\begin{equation}\label{eq:biascor}
    b_o^\prime =  b_o + \sum_{i=1}^N \mathrm{E}[\epsilon_{W_{o,i}}]
\end{equation}
where $\mathrm{E}[\epsilon_{W_{o,i}}]$ is the mean error of the approximate multiplication $W_{o,i} \times X_i$, $\forall X_i$.
Hence, the mean convolution error is given by~\cite{Tasoulas2020:weightoriented}:
\begin{equation}\label{eq:convzero}
\begin{split}
\mathrm{E}[\epsilon_{Y_o}] & = \mathrm{E}[Y_o-Y_o^\prime] \\
& = b_o-b_o^\prime- \sum_{i=1}^N \mathrm{E}[\epsilon_{W_{o,i}}] = 0
\end{split}
\end{equation}
% As a result, given an approximate multiplier, by just updating the bias term using~\eqref{eq:biascor}, the mean convolution error is nullified.
\blue{As a result, by just updating the bias term using~\eqref{eq:biascor} the mean convolution error is effectively nullified.\label{commentR3C13}
However, fully exact inference cannot be achieved since the convolution error features non-zero variance.
To demonstrate the impact of the bias update,~\cite{Tasoulas2020:weightoriented} showed that for the same accuracy loss constraints, the bias update enables using higher approximation and more energy-efficient multipliers.
For example, for $0.5$\% accuracy loss constraint, using the bias update~\cite{Tasoulas2020:weightoriented} achieved $1.4$x higher energy reduction compared to the case that does not consider the bias update.
% Though, the direct effect of the bias update on DNNs accuracy is not mathematically analyzed in~\cite{Tasoulas2020:weightoriented}.
}

Finally, Zervakis et al.~\cite{Zervakis2021:ControlVar} introduced a control variate approximation to improve the accuracy of the convolution operation.
Instead of~\eqref{Neuron_Equation},~\cite{Zervakis2021:ControlVar} proposed to compute:
\begin{equation}\label{eq:convar}
    Y_o = b_o + \sum_{i=1}^{N} M_{ax}(W_{o,i},X_i) + V_o
\end{equation}
again $M_{ax}$ corresponds to the approximate multiplication and $V_o$ is the proposed control variate.
The selection of $V_o$ depends on the approximate multiplier.
The control variate $V_o$ estimates the convolution error at runtime and though an extra addition, $V_o$ is added to the approximate convolution result to mitigate the error.
It is mandatory that $V_o$ can be easily computed in order not to annihilate the gains of the employed approximation ($M_{ax}$).
In~\cite{Zervakis2021:ControlVar}, $V_o$ is calculated as a function of the input activations and the average value of the weights.
Specifically, in~\cite{Zervakis2021:ControlVar}, $V_o$ is given by:
\begin{equation}\label{eq:vo}
    V_o = \overline{W} \sum_{i=1}^{N}(X_i \text{ mod } 2^m),\quad
    \overline{W}=\frac{1}{N}\sum_{i=1}^{N}W_{o,i}, 
\end{equation}
where $m$ is a configuration parameter of the perforated approximate multiplier~\cite{Zervakis:TVLSI2016:perf} that used in~\cite{Zervakis2021:ControlVar}.
Higher $m$ refers to higher approximation and higher energy gains.
\cite{Zervakis2021:ControlVar} demonstrated that the proposed control variate approximation technique nullifies the mean convolution error and minimizes its variance.
\blue{Over VGG-13/16, ResNet-44/56, ShuffleNet, and GoogleNet trained on CIFAR10, \cite{Zervakis2021:ControlVar} improved, the inference accuracy, on average, from $0.86$\%  (when $m$=$1$) up to $21$\% (when $m$=$3$) compared to the same approximation without the control variate (i.e., using~\cite{Zervakis:TVLSI2016:perf} in~\eqref{eq:convar} without $V_o$).\label{commentR3C14}
For the same CNNs on CIFAR100, the respective improvement is from $3.6\%$ to $21\%$.}

\subsection{Error Metric Optimization}\label{subsec:errormetric}

Many hardware approximation algorithms and frameworks are usually guided by the mean relative error distance (MRED) metric~\cite{errormetric,Zervakis:ACCESS2020}.
Nevertheless, MRED might not be an optimal metric for approximate DNN inference accelerators.
To increase the achieved accuracy, several works optimize the generated approximate multipliers targeting specific error metrics.
The generated approximate multipliers are most suitable for DNNs and can achieve higher accuracy when combined with the previously analyzed compensation techniques or even when applied in isolation.
In~\cite{Mrazek2016:CGPevo} the authors design approximate multipliers that satisfy:
\begin{equation}
\begin{gathered}
|M_{ax}(w,a)-w\cdot a| \leq c\,\forall a,\,\forall w\text{ and }\\
M_{ax}(0,a)=M_{ax}(w,0)=0\,\forall a,\,\forall w
\end{gathered}
\end{equation}
again $M_{ax}$ corresponds to the approximate multiplication and $c$ is an error threshold.
In other words,~\cite{Mrazek2016:CGPevo} ensures that the worst-case error of the approximate multiplier is below a given threshold and that multiplication by $0$ is always accurate.
Then,~\cite{Mrazek2016:CGPevo} performs an exploration to find the maximum value of $c$ that a given DNN can tolerate.
Mrazek et al.~\cite{Mrazek2016:CGPevo} concluded that although the impact of approximate multipliers on the accuracy is DNN-specific, aiming for high accuracy it is mandatory to have the accurate multiplication by $0$.
The same error metric is used in~\cite{Mrazek2020:Libraries}.
Vasicek et al.~\cite{Vasicek2019:DataDistrib} used the weighted mean error distance (WMED), which considers the input data distribution:
\begin{equation}\label{eq:wmed}
\frac{1}{|\{a|\forall a\}||\{w|\forall w\}|}\sum_{\forall a}\sum_{\forall w} \mathrm{D}(w)|M_{ax}(w,a)-w\cdot a| \leq c,
\end{equation}
where $\mathrm{D}$ is the probability mass function and $c$ is an error threshold.
Using~\eqref{eq:wmed} and exploiting that the weights are known after training (and thus $\mathrm{D}$),~\cite{Vasicek2019:DataDistrib} assigns higher significance to the weights that appear more often (i.e., higher $\mathrm{D}(w)$).
Hence,~\cite{Vasicek2019:DataDistrib} tries to ensure that the more often multiplications are performed more accurately, leading to higher inference accuracy overall.
Ansari et al.~\cite{Ansari2020:ImprovEvo} evaluated several error metrics in order to identify critical features that render an approximate multiplier suitable for DNN inference.
Specifically, the error rate (ER), the error distance (ED), the absolute ED (AED) and the relative ED (RED) metrics were examined.
Using these error metrics, nine relevant error features of the approximate multipliers were evaluated.
These features are reported in Table~\ref{tab:errorfeature}.
Through extensive experimentation,~\cite{Ansari2020:ImprovEvo} concluded that the most important features that make an approximate multiplier superior to others are $\mathrm{Var}$(ED) and $\mathrm{RMS}$(ED).
Tasoulas et al.~\cite{Tasoulas2020:weightoriented} reached the same conclusion.
Through a rigorous mathematical analysis,~\cite{Tasoulas2020:weightoriented} demonstrated that the mean convolution error can be cancelled using a constant correction term as~\eqref{eq:biascor}-\eqref{eq:convzero} show.
Thus,~\cite{Tasoulas2020:weightoriented} deduced that the ED variance ($\mathrm{Var}$(ED)) is a more important error feature when designing approximate multipliers for DNN inference.
\cite{Tasoulas2020:weightoriented} showed that the mean and variance values of $\epsilon_{Y_o}$, i.e., of the convolution error (ED), are given by:
\begin{equation}\label{eq:muvarconv}
\begin{gathered}
\mathrm{E}[\epsilon_{Y_o}]  = 0 \text { and }\\
\mathrm{Var}(\epsilon_{Y_o}) = \sum_{i=1}^{N}\mathrm{Var}(\epsilon)=N\mathrm{Var}(\epsilon),
\end{gathered}
\end{equation}
where $ Var(\epsilon)$ is the error (ED) variance of the approximate multiplier \blue{and N is the number of the filter’s weights.} 
Note that the variance after the constant compensation (i.e., \eqref{eq:biascor}) equals the mean squared error (i.e., square of $\mathrm{RMS}$(ED)).
Hence, the mathematical analysis of~\cite{Tasoulas2020:weightoriented} is validated by the experimental findings of~\cite{Ansari2020:ImprovEvo} and vice versa.

\begin{table}[t!]
\renewcommand{\arraystretch}{1.1}
\caption{Error features evaluated in \cite{Ansari2020:ImprovEvo}.}
\label{tab:errorfeature}
\centering
\footnotesize
\vspace{-0.3cm}
\begin{tabular}[t]{l|l}
\hline
\textbf{Feature} & \textbf{Description} \\ \hline
ER & Error Rate \\
$\mathrm{Var}$(ED) & Variance of ED \\
$\mathrm{E}$[ED] & Mean value of ED \\
$\mathrm{RMS}$(ED) & Root Mean Square of ED values \\
$\mathrm{Var}$(RED) & Variance of RED \\
$\mathrm{E}$[RED] & Mean value of RED \\
$\mathrm{RMS}$(RED) & Root Mean Square of RED values \\
$\mathrm{Var}$(AED) & Variance of AED \\
$\mathrm{E}$[AED] & Mean value of AED \\\hline
\end{tabular}

\end{table}
\section{Energy-Accuracy Evaluation}\label{sec:evaluation}

In this section we evaluate the efficiency, in terms of potential energy reduction and accuracy loss, of hardware approximation when targeting CNN accelerators.
As we will observe in the remainder of this analysis, both the accuracy loss and the energy savings depend on the complexity of the evaluated use cases/benchmarks (i.e., neural network, dataset, and/or precision).
Hence, we first provide a comprehensive discussion regarding the complexity of the considered use cases and then we evaluate the accuracy-energy results as obtained from the respective publications of Section~\ref{sec:techniques}.

\subsection{Assessing the Complexity of the Evaluation Scenarios}\label{subsec:datasets}

To assess the efficiency of approximate DNN accelerators, it is mandatory to analyze the datasets that are used to evaluate the accuracy loss due to the introduced approximation.
In other words, to evaluate how efficient an approximation is, we need to determine the complexity of the benchmark that was used to measure the attained accuracy after the approximation.
For example, although the MNIST dataset is widely used in early ML research, it is a fairly simple dataset and it is fairly easy to achieve high accuracy even with high approximation~\cite{Mrazek2016:CGPevo}.
As a result, MNIST cannot be considered as a representative example of the complex services that modern DNN-based systems deliver today.
Hence, impressing results observed for the MNIST dataset, are hardly expected to be achieved in more complex datasets.

\begin{figure}[t]
 \centering
\includegraphics[width=1\linewidth]{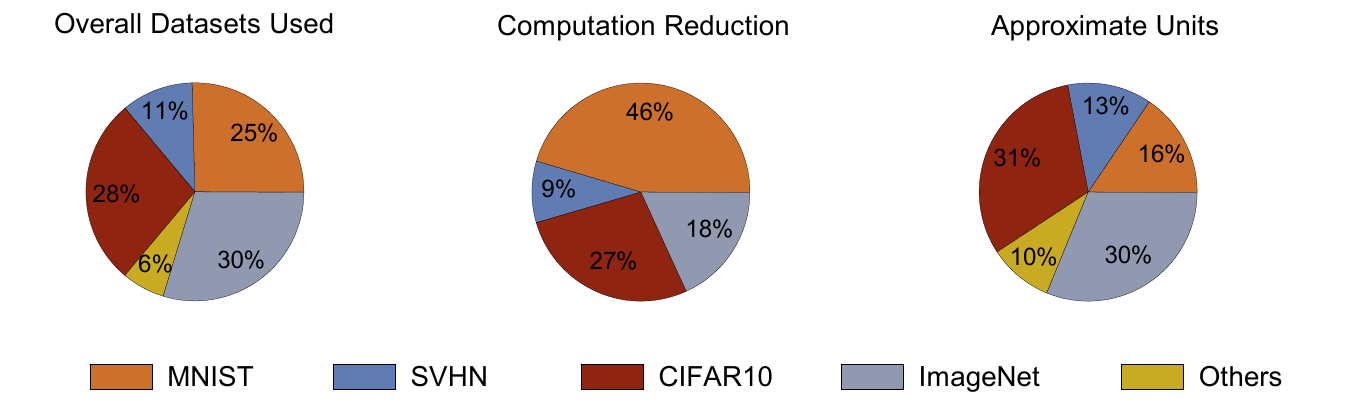}%
\centering
\caption{
The allocation of the datasets that are used in the evaluation of hardware DNN approximation.
The pies present the percentage of the accuracy/energy evaluations that examine the respective dataset.
The middle and right pies refer to the Computation Reduction and Approximate Units categories, while the left pies presents the aggregated results among all the approximate works.
This figure evaluates the complexity of the performed evaluations with respect to the dataset difficulty.
\blue{To generate this figure, all the works described in Section~\ref{subsec:compred} and~\ref{subsec:axunits} have been considered.}
}
\label{fig:pites}%
\end{figure}
\begin{figure}[t]
 \centering
\includegraphics[width=1\linewidth]{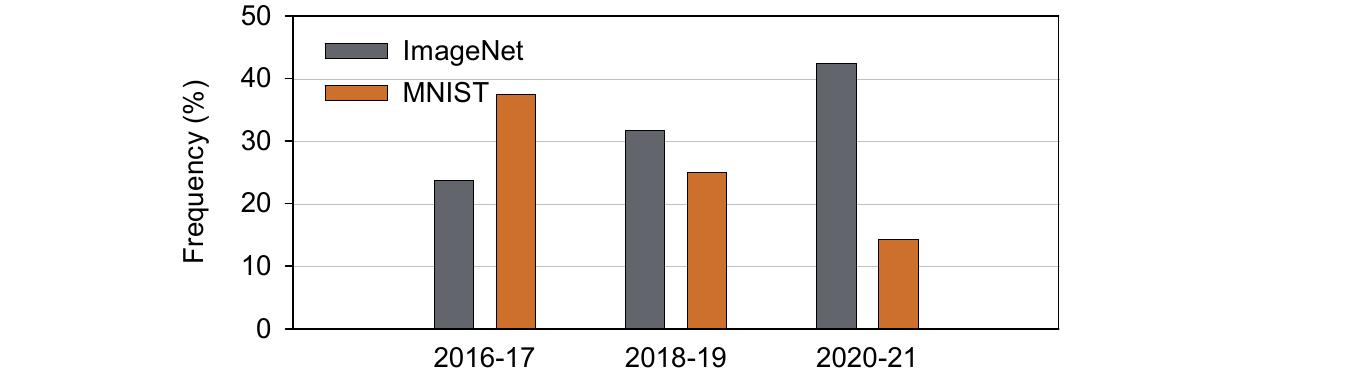}%
\centering
\caption{
Evolution of the number of approximate works (in percentage) that target MNIST and ImageNet.
\blue{All the works of Sections~\ref{subsec:compred} and~\ref{subsec:axunits} are considered to generate this figure.}
}
\label{fig:Im_vs_MN}%
\end{figure}

%\blue{Among all the works presented and analyzed in Sections~\ref{subsec:compred} and~\ref{subsec:axunits}, various datasets have been used to evaluate the accuracy of the proposed DNN approximation techniques.}\label{commentR3C16} 
Among all the works analyzed in our survey \blue{(techniques discussed in Sections~\ref{subsec:compred} and~\ref{subsec:axunits})}\label{commentR3C16}, various datasets have been used to evaluate the accuracy of the proposed DNN approximation techniques.
In Fig.~\ref{fig:pites}, we present the respective dataset allocation.
As shown, the most popular datasets are MNIST, SVHN, CIFAR10 and ImageNet.
Overall, there is a quite balanced research effort distributed among the simple MNIST dataset and the more complex CIFAR10 and ImageNet.
It is noteworthy that the Computation Reduction approximation category mainly targets MNIST while the Approximate Units category focuses on CIFAR10 and ImageNet.

Despite the fact that several works targeted the fairly simple MNIST dataset, Fig.~\ref{fig:Im_vs_MN} demonstrates that DNN hardware approximation follows the current research trend.
The evolution of DNNs has led researchers to target more complex use cases.
Over time, as shown in Fig.~\ref{fig:Im_vs_MN}, more and more DNN hardware approximation works focus on the ImageNet dataset while fewer and fewer works target MNIST.
For example, the percentage of approximate DNN techniques that used ImageNet increased from 24\% in 2016/17 to 42\% in 2020/21.
In the same period the percentage of works that use MNIST decreased from 38\% to 14\%.
\begin{figure}[t]
 \centering
\includegraphics[width=1\linewidth]{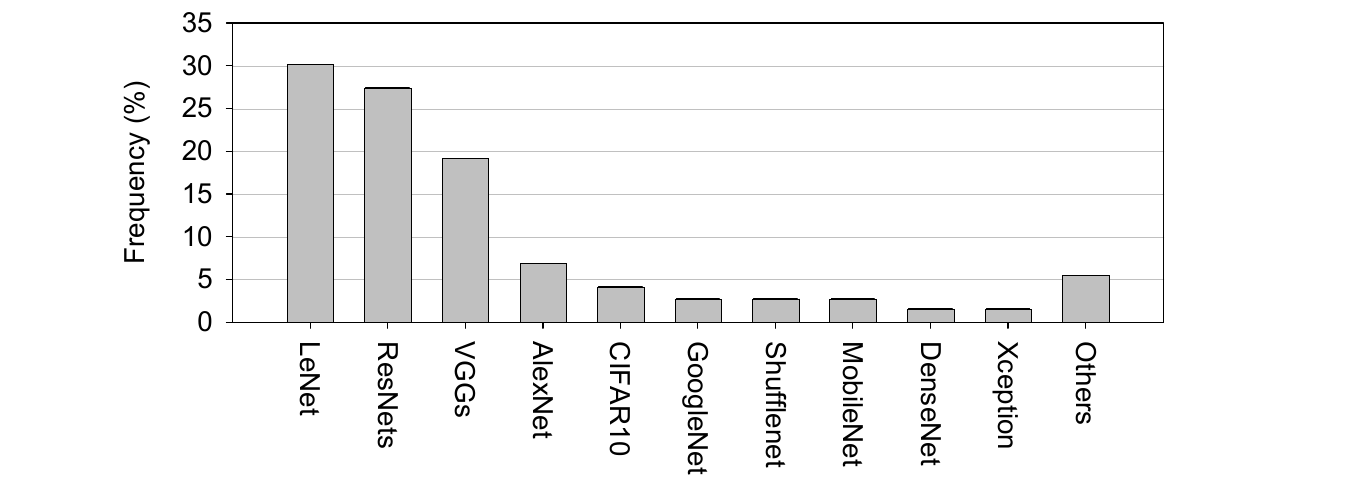}%
\centering
\caption{
The DNNs that are examined in the accuracy/energy evaluations of the approximate works.
Each bar represents the number of works (in percentage) that used the respective DNN.
}
\label{fig:Networks_percent}%
\end{figure}
\begin{figure}[t]
 \centering
\includegraphics[width=1\linewidth]{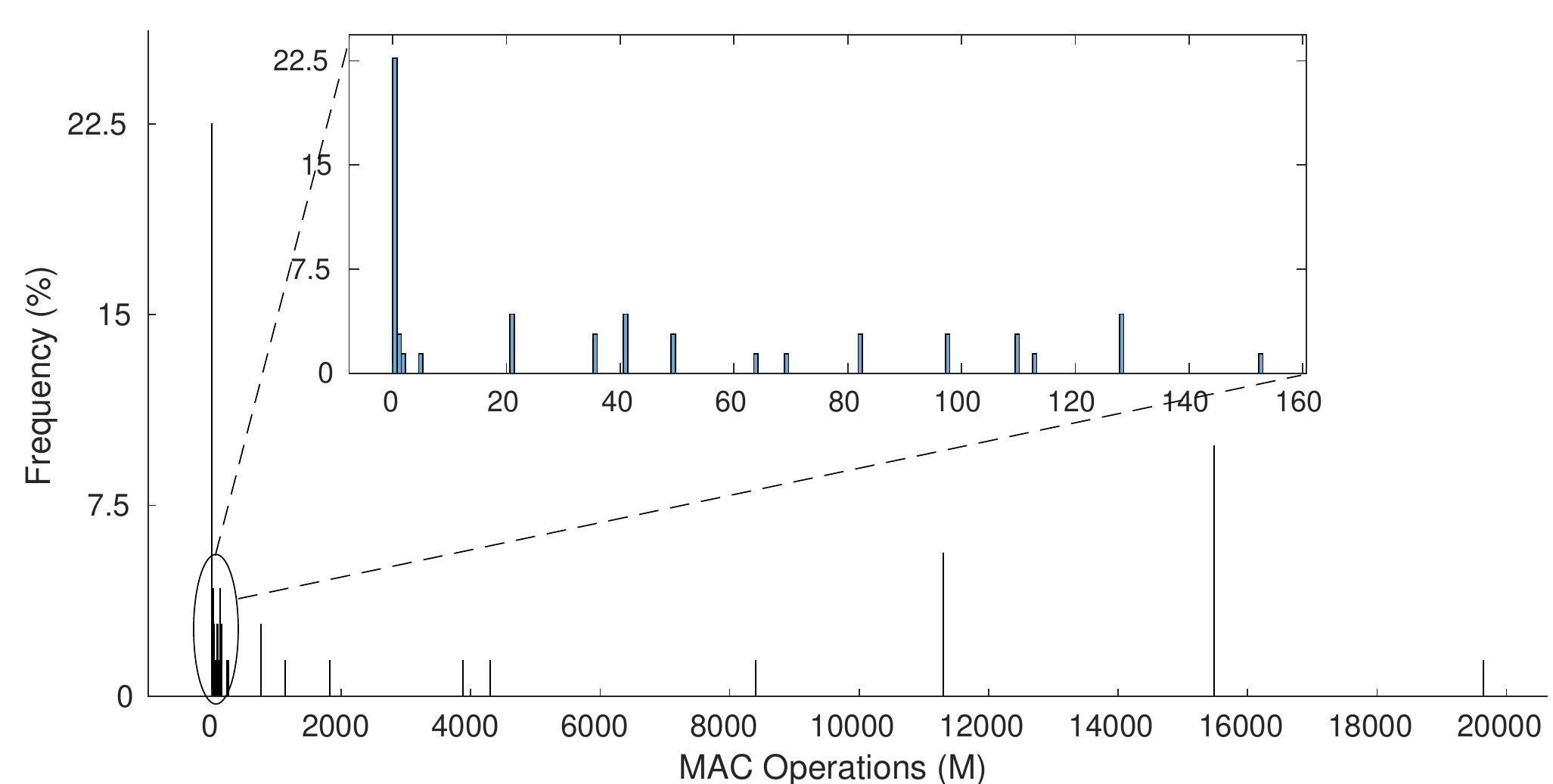}%
\centering
\caption{
The number of Mac Operations required by the DNNs that are used in the accuracy/energy evaluations of the approximate works.
Each bar represents the number of works (in percentage) that considered a DNN of the respective size.
This figure evaluates the complexity of the performed evaluations w.r.t. the DNN size.
}
\label{fig:histogram_flops}%
\end{figure}

Moreover, in addition to the considered dataset, it is also essential to examine the complexity of the DNNs that are used to evaluate the accuracy of the DNN approximation techniques.
In Fig.~\ref{fig:Networks_percent}, we present the most widely used DNNs evaluated on our survey.
As shown, the $30$\% of the accuracy evaluations are performed on the fairly simple LeNet network.
Nonetheless, significant research also targets complex networks such as the VGG (19\%) and ResNet (27\%) networks.
Fig.~\ref{fig:histogram_flops} presents a more descriptive view of Fig.~\ref{fig:Networks_percent}.
In Fig.~\ref{fig:histogram_flops}, we analyze the size (in terms of numbers of required MAC operations) of the DNNs used in the accuracy evaluation of the state-of-the-art approximation techniques.
Note that the numbers of MAC operations depends on both the number of the DNN parameters as well as the input size (i.e., dataset used).
As shown in Fig.~\ref{fig:histogram_flops}, many approximate works (i.e., the 44\%) examine DNNs that require only a few tens of million ($<100$M) MAC operations.
Nevertheless, significant research ($>24\%$) is also performed on larger DNNs that require billions  ($>2$G) of MAC operations.

It is noteworthy that although the main objective of hardware approximation is energy efficiency (that is crucial especially for embedded devices) only a small portion of DNN approximation techniques targeted mobile-oriented DNNs such as MobileNet and Squeezenet (included in others).
\blue{Although this might be explained by the fact that such networks are already very compressed becoming, thus,
very sensitive to further approximation, a wider and more comprehensive evaluation (i.e., more approximation techniques must be evaluated on such challenging architectures~\cite{Jacob2018:Quant}) is required to draw such conclusions.}\label{commentR3C18}

Finally, the complexity of the evaluation is highly correlated to the precision that is used to represent the weights and activations.
Low precision representations require low bitwidth arithmetic and thus smaller circuits (e.g., multipliers and adders).
As a result, they constitute more challenging use cases to apply approximate computing since they exhibit limited space for additional approximation.
In other words, compared to $8$-bit implementations, it is easier to achieve high energy savings combined with small accuracy loss when $32$ bits are used for weights and activations.
However, note that $8$-bit precision is mainly used today in the state-of-the-art exact DNN accelerators~\cite{google:tpu}.

\begin{figure}[t]
 \centering
\includegraphics[width=0.92\linewidth,height=4.5cm]{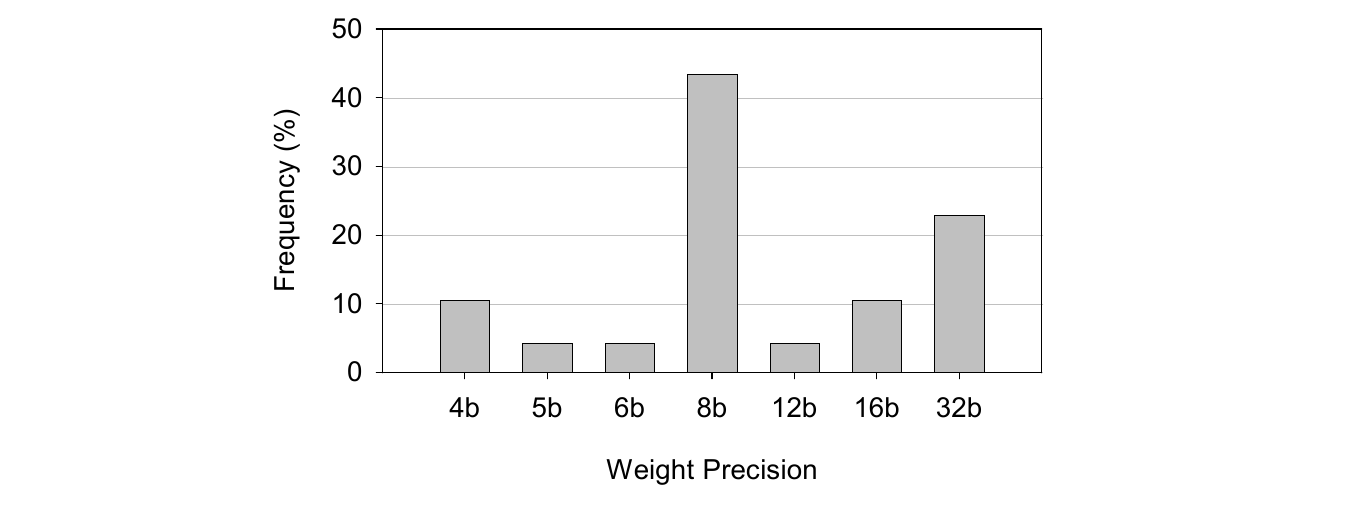}%
\centering
\vspace{-0.5cm}
\caption{
The weight precision that is used in the accuracy/energy evaluations of the approximate works.
Each bar represents the number of works (in percentage) that used the respective precision.
This figure evaluates the complexity of the performed evaluations with respect to the inference precision.
}
\label{fig:precision_distrib}%
\end{figure}

%\blue{
Fig.~\ref{fig:precision_distrib} presents the weight precision that is used in the evaluation of the approximate DNN techniques.
Similar results are obtained for the precision of the activations.
%Weights are more of importance in arithmetic analysis and number formats than activations, as they can be already known in an approximate trained DNN.
As shown in Fig.~\ref{fig:precision_distrib}, a considerable amount of the approximation techniques  ($>30\%$) uses high precision (i.e., $\geq16$b).
%This percentage mainly concerns Memoization techniques where high bit-width arithmetic is of a vital meaning than other families, which is also confirmed from the Table~\ref{tab:compred}.
%The rest percentage is distributed among weight values represented with equal or lower than 8b.
%The high percentage of 8b precision (>40\%) confirms the trend of low bit-width approximate techniques becoming more and more widely used.
%The remarkable thing of such designs is that although they are basically present lower energy gains, their energy consumption is also low.
Nevertheless, the majority of the works ($>40$\%) examine the conventional $8$b precision.
It is noteworthy that many works ($>18$\%) examined also more challenging cases in which very low precision is used for the weights ($\leq6$b).

\subsection{Performance Analysis}\label{subsec:performanceanal}

In this subsection we evaluate the energy reduction and accuracy loss achieved by the CNN hardware approximation techniques that are analyzed in Sections~\ref{subsec:compred}-\ref{subsec:axunits}.
%\blue{It is noteworthy that most of our related state-of-the-art works are mainly focused on CNNs and hence, the following analysis concerns only this type of networks.}\label{commentR1C1c}
Tables~\ref{tab:compred}-\ref{tab:axun_1} present the corresponding results.
For each technique, the columns Neural Network and Dataset present the neural network model and dataset that was used in the respective evaluation.
The columns \#Conv Layers and \#MAC Ops report the number of convolutional layers and the number of MAC operations required for the corresponding neural network.
\blue{The required MAC operations of~\cite{Tasoulas2020:weightoriented, Hammad2021:controller,Mrazek2019:ALWANN,Mrazek2020:Libraries} are calculated using the data reported in the corresponding paper while for the rest works we used pytorchcv~\cite{pytorchWebsite}.\label{commentR3C20}
Though, some works don't provide adequate information to calculate the MAC operations of their employed networks.}
The column Precision Baseline refers to the precision that the exact (baseline design) uses for the weights and activations.
A \textit{x}-bit baseline uses \textit{x} bits to represent the weights and activations as well as a \textit{x}-bit exact multiplier to perform the multiplications.
The column Precision Approximate refers to the precision that the approximate implementation uses for the weights and activations.
The column Acc Loss/Energy Sav. presents the energy reduction and accuracy loss of the approximate implementation with respect to the corresponding baseline.

\begin{table}[]
\caption{Evaluation of Computation Reduction Approximation Category }
\label{tab:compred}

\begin{threeparttable}
\footnotesize
\renewcommand{\arraystretch}{1.1}
\begin{tabular}{cccccccc}
\hline
\textbf{Ref}   & \makecell{\textbf{Neural}\\ \textbf{Network} } & \textbf{Dataset} & \makecell{\textbf{Precision}\\ \textbf{Approximate} } & \makecell{ \textbf{Precision}\\ \textbf{Baseline} } & \makecell{ \textbf{\#Conv}\\ \textbf{Layers} } & \textbf{\#MAC Ops} & \makecell{ \textbf{Acc. Loss(\%)/}\\ \textbf{Energy Sav.(\%)} } \\ \hline \hline
\multicolumn{8}{c}{\textbf{Memoization}}                                                                                                                                                                                       \\ \hline
\multirow{3}{*}{\cite{Mocerino2019:CompReuseFP}}       & LeNet          & MNIST            & 32b                & 32b               & 2                      & 0.34M               & 0.50 / 69                                                                   \\
               & LeNet          & GTRSB            & 32b                & 32b               & 2                      & 0.34M               & 0.10 / 22                                                                   \\
               & Custom         & GSC              & 32b                & 32b               & -                      & -                 & 2.00 / 57                                                                     \\ \hline
{\cite{Jiao2018:BloomFilter}}        & LeNet          & MNIST            & 32b                & 32b               & 2                      & 0.34M               & 0.50 / 45                                                                 \\ \hline

\multirow{3}{*}{\cite{InpSimilarity:2018}}       & Kaldi          & Librispeech      & 32b                & 32b               &   0                     & 4.64M                & 0.47 / 49                                                                  \\
               & C3D            & UCF101           & 32b                & 32b               &    8                    & 0.11G             & 1.38 / 77                                                                  \\
               & Autopilot      & Videos           & 32b                & 32b               &  5                      & 2.22M              & 0.06 / 76                                                                  \\ \hline \hline
\multicolumn{8}{c}{\textbf{Skipping}}                                                                                                                                                                                          \\ \hline
{\cite{Kim2021:PSandZS}}       & VGG-16         & ImageNet         & 12b                & 12b               & 13                     & 15.48G              & 0.80 / -\tnote{1}                                                                \\ \hline
\multirow{5}{*}{\cite{Hemmat2020a:AirNN}}       & VGG-16         & ImageNet         & 32b                & 32b               & 13                     & 15.48G              & 1.43 / 24                                                                  \\
               & MobileNetV2   & ImageNet         & 32b                & 32b               &  35                      & 0.33G              & 2.70 / 15                                                                    \\
               & LeNet300-100  & MNIST            & 32b                & 32b               & 0                      & 0.27M             & 0.12 / 89                                                                  \\
               & LeNet       & MNIST            & 32b                & 32b               & 2                      & 0.34M               & 1.29 / 51                                                                  \\
               & CIFAR10        & CIFAR10          & 32b                & 32b               &  -                      & -              & 2.00 / 68                                                                     \\ \hline
\multirow{4}{*}{\cite{Piyasena2019:PredictandSkip}}       & VGG-16         & ImageNet         & 8b                 & 8b                & 13                     & 15.48G              & 0.21 / 10                                                               \\
               & AlexNet        & ImageNet         & 8b                 & 8b                & 5                      & 1.13G               & 0.27 / 10                                                               \\
               & CIFAR10-Quick  & CIFAR10          & 8b                 & 8b                &      3                  & 0.33M              & 0.45 / 12                                                               \\
               & LeNet          & MNIST            & 8b                 & 8b                & 2                      & 0.34M               & -0.02 / 10                                                              \\ \hline
\multirow{2}{*}{\cite{Ujiie2016:SignPredictandSkip}}       & Custom         & MNIST            & 32b                & 32b               & 2                      & -               & -0.20 / 14                                                                \\
               & Custom         & CIFAR10         & 32b                & 32b               & 2                      & -               & 1.10 / 14                                                                 \\ \hline
{\cite{Huan2016:NearZero}} & Custom         & MNIST            & 16b                & 16b               &  -                      & 0.08M                 & \textless 1.00 / 74      \\ \hline                                                  
\end{tabular}
\begin{tablenotes}
\item[1]\cite{Kim2021:PSandZS} reports $1.7$x higher energy-efficiency (TOPS/W).
\end{tablenotes}
\end{threeparttable}%

\end{table}

Table~\ref{tab:compred} presents the evaluation of the Computation Reduction category and its Memoization and Skipping subcategories.
As shown in Table~\ref{tab:compred}, the Memoization approximation family achieves very high energy reduction (up to $77$\%) for a minimal accuracy loss ($0.53$\%) on average.
Nevertheless, only $32$-bit precision is considered in this family and the examined networks are shallow (up to $8$ convolutional layers).
On the other hand, the Skipping family examines deeper networks, e.g., VGG-16 and MobileNet with $13$ and $35$ convolution layers, respectively.
In addition, the DNNs examined from the Skipping family feature many MAC operations ($4.8$G on average).
With respect to the employed precision, the Skipping approximation techniques mainly use high precision (i.e., $16$-bit and $32$-bit precision).
However, \cite{Kim2021:PSandZS} and~\cite{Piyasena2019:PredictandSkip} use lower precision, i.e., $12$-bit and $8$-bit respectively.
Again, when considering simpler evaluation cases (e.g., LeNet, MNIST, and/or $32$-bit precision) Skipping approximation delivers minimal accuracy loss and very high energy gains.
On the contrary, this is not the case for more complex evaluations.
When considering $8$-bit precision the energy gains drop to around $10$\%.
Similarly, when ImageNet is considered (even with $32$-bit precision) the energy savings drop significantly and the accuracy loss increases.
For example,~\cite{Hemmat2020a:AirNN} achieved $24$\% energy reduction and $1.43$\% accuracy loss on VGG-16 while the respective values are $15$\% and $2.7$\% when considering MobileNet.
Notably, with $12$-bit precision, \cite{Kim2021:PSandZS} features $1.7$x better energy efficiency (TOPS/W) and only $0.8$\% accuracy loss for VGG-16 on the ImageNet dataset.

Tables~\ref{tab:axun_2}-\ref{tab:axun_1} present the respective analysis for the Approximate Units category.
The results for the Multiplierless and Log-Multipliers families are reported in Table~\ref{tab:axun_2}.
As shown in Table~\ref{tab:axun_2}, the Multipliers approximation targets low precision inference.
This family exhibits a broad evaluation that covers a wide spectrum from simple use cases to very complex ones (such as ResNet-164 on CIFAR100 and ResNet-50 on ImageNet).
Remarkably, with $8$-bit precision, \cite{Sarwar2018:Alphabet} achieved $53$\% energy reduction (compared to the $12$-bit baseline) and only $0.37$\% accuracy loss for the very challenging ResNet-164 on CIFAR100.
Note that although~\cite{Sarwar2018:Alphabet} used only $4$-bit for simpler benchmarks (e.g., MNIST, SVHN) it required $8$-bit precision for more complex evaluations (e.g., CIFAR10 and CIFAR100).
As Table~\ref{tab:axun_2} also illustrates, the Approximate Log-Multipliers family focuses on more complex datesets such as CIFAR10 and ImageNet and examines varying precision values ($4$-bit to $32$-bit).
As in the previous techniques, when $32$-bit is used, very high energy reduction is achieved (more than $70$\%) with a negligible accuracy loss (less than $0.5$\%).
On the other hand, when the employed precision decreases, the energy savings drop significantly (down to $22$\% for $4$ bits) and the accuracy loss increases (to $4.32$\% for $4$ bits).
Still, the obtained energy savings are considerable.

\begin{table}[]
\caption{Evaluation of Approximate Units Approximation Category (Multiplierless \& Log-Multipliers)}
\label{tab:axun_2}
\begin{threeparttable}
\footnotesize
\renewcommand{\arraystretch}{1.1}
\begin{tabular}{cccccccc}
\hline
\textbf{Ref}   & \makecell{\textbf{Neural}\\ \textbf{Network} } & \textbf{Dataset} & \makecell{ \textbf{Precision}\\ \textbf{Approx. (W/A)} } & \makecell{ \textbf{Precision}\\ \textbf{Baseline} } & \makecell{ \textbf{\#Conv}\\ \textbf{Layers} } & \textbf{\#MAC Ops} & \makecell{ \textbf{Acc. Loss(\%)/}\\ \textbf{Energy Sav.(\%)} } \\ \hline \hline
\multicolumn{8}{c}{\textbf{Multiplierless}}                                                                                                                                                                                  \\ \hline
\multirow{5}{*}{\cite{Sarwar2018:Alphabet}$^*$}      & MLP            & MNIST            & 4b                 & 12b               & 0                      & 0.08M                 & 0.35 / 61                                                               \\
             & MLP            & TiCH             & 4b                 & 12b               & 0                      &   0.42M               & 1.71 / 79                                                               \\
             & MLP            & SVHN             & 4b                 & 12b               & 0                      & 1.05M                 & 1.68 / 79                                                               \\
             & NIN            & CIFAR10          & 8b                 & 12b               & 2                      & 0.97M             & 0.00 / 53                                                                  \\
             & ResNet-164(BN)     & CIFAR100         & 8b                 & 12b               & 163                       & 0.26G              & 0.37 / 53                                                               \\ \hline
\multirow{3}{*}{\cite{Faraone2020:AddNet}$^*$}     & AlexNet        & ImageNet         & 4b / 8b             & 8b                & 5                      & 1.13G               & -0.30 / 25                                                                     \\
             & ResNet-18      & ImageNet         & 4b / 8b              & 8b                & 17                     & 1.82G               & 0.90 / -\tnote{1}                                                                       \\
             & ResNet-50      & ImageNet         & 4b / 8b              & 8b                & 49                     & 3.88G            & 0.40 / -\tnote{1}                                                                       \\ \hline
{\cite{Parmar2020:CORDIC}}     & VGG-16         & ImageNet          & 8b                 & 8b                & 13                     & 15.48G              & 0.10 / 55\tnote{2}          \\ \hline    \hline   
\multicolumn{8}{c}{\textbf{Approximate Log-Multipliers}}                                                                                                                                                                                 \\ \hline
\multirow{2}{*}{\cite{TwoStageLog:2021}$^*$}     & ResNet-20      & CIFAR10         & 8b                 & 32b                & 19                     & 41.29M             & 3.14 / 53                                                               \\
             & ResNet-20      & CIFAR10         & 16b                & 32b               & 19                     & 41.29M             & 0.50 / 76                                                                \\ \hline
\multirow{2}{*}{\cite{Kim2019:mitchell}}     & Cuda-convnet   & CIFAR10          & 32b                & 32b               & 3                      & 0.33M                 & 0.00 / 74                                                                     \\
             & AlexNet        & ImageNet         & 32b                & 32b               & 5                      & 1.13G               & 0.30 / 74                                                                   \\ \hline
\multirow{2}{*}{\cite{Vogel2018:ArbitraryLog}}     & VGG-16         & ImageNet          & 5b                 & 8b                & 13                     & 15.48G              & 2.72 / 22\tnote{3}                                                                \\
             & AllCNN         & ImageNet         & 4b                 & 8b                & 9                      & -              & 4.32 / 22\tnote{3}                                                                \\ \hline
{\cite{Parames2018:FPMult}}     & AlexNet        & ImageNet         & 32b                & 32b               & 5                      & 1.13G               & -1.00 / 73\tnote{3}                                                                 \\ \hline                                                
\end{tabular}
\begin{tablenotes}
\item[1] \cite{Faraone2020:AddNet} reports only area reduction (up to 55\% LUTs reduction). %for this network
\item[2] \cite{Parmar2020:CORDIC} is compared only to recent proposed architectures and reports about 55\% power savings.
\item[3] The same operating frequency is assumed for the approximate and baseline designs
\item[*] \blue{Retraining/Fine-tuning is used (see Section~\ref{subsec:retraining})}
\end{tablenotes}
\end{threeparttable}%
\end{table}

%%%%%%%%%%%%%%%%%%%%%%%%%%%%%%%%%%%%%%%%
%%%%%%%%%%%%%%%%%%%%%%%%%%%%%%%%%%%%%%%%

\begin{table}[]
\caption{Evaluation of Approximate Units Approximation Category (Approximate Multipliers/Adders)}
\label{tab:axun_1}
\begin{threeparttable}
\footnotesize
\renewcommand{\arraystretch}{1.1}
\vspace{-2mm}
\begin{tabular}{cccccccc}
\hline
\textbf{Ref}   & \makecell{\textbf{Neural}\\ \textbf{Network} } & \textbf{Dataset} & \makecell{ \textbf{Precision}\\ \textbf{Approx. (W/A)} } & \makecell{ \textbf{Precision}\\ \textbf{Baseline} } & \makecell{ \textbf{\#Conv}\\ \textbf{Layers} } & \textbf{\#MAC Ops} & \makecell{ \textbf{Acc. Loss(\%)/}\\ \textbf{Energy Sav.(\%)} } \\ \hline \hline
\multicolumn{8}{c}{\textbf{Approximate Multipliers/Adders}}                                                                                                                                                                         \\ \hline
\multirow{4}{*}{\cite{Mrazek2016:CGPevo}$^*$}      & LeNet        & MNIST            & 8b                 & 8b                & 2                      & 0.34M               & 0.09 / 77                                                               \\
             & LeNet        & SVHN             & 8b                 & 8b                & 2                      & 0.34M               & -0.07 / 57                                                               \\
             & LeNet        & MNIST            & 12b                & 12b               & 2                      & 0.34M               & -0.02 / 66                                                              \\
             & LeNet        & SVHN             & 12b                & 12b               & 2                      & 0.34M               & -0.01 / 66                                                              \\ \hline
\multirow{2}{*}{\cite{Ansari2020:ImprovEvo}$^*$}      & MLP            & MNIST            & 8b                 & 8b                & 0                      & 0.24M                 & -0.01 / 71                                                              \\
             & LeNet        & SVHN             & 8b                 & 8b                & 2                      & 0.34M               & -0.07 / 71                                                              \\ \hline
\multirow{4}{*}{\cite{Mrazek2020:Libraries}$^*$}     & ResNet-8       & CIFAR10          & 6b / 8b            & 8b                & 7                      & 21.10M              & 0.32 / 37                                                                \\
             & ResNet-14      & CIFAR10          & 6b / 8b            & 8b                & 13                     & 35.30M               & 0.18 / 37                                                                \\
             & ResNet-20      & CIFAR10          & 5b / 8b            & 8b                & 18                     & 49.50M             & 0.34 / 56                                                                \\
             & ResNet-26      & CIFAR10          & 4b / 8b            & 8b                & 25                     & 63.60M               & 1.21 / 70                                                               \\ \hline
\multirow{2}{*}{\cite{Vasicek2019:DataDistrib}}     & MLP            & MNIST            & 8b                 & 8b                & -                      & 0.24M                 & -0.20 / 60                                                                  \\
             & LeNet        & SVHN             & 8b                 & 8b                & 2                      & 0.34M               & -0.12 / 70                                                                 \\ \hline
{\cite{Hanif2019:CANN}}     & LeNet        & CIFAR10          & 8b                 & 8b                & 2                      & 0.34M               & -1.08 / 46                                                              \\ \hline
\multirow{3}{*}{\cite{Mrazek2019:ALWANN}$^\dagger$}     & ResNet-8       & CIFAR10          & 8b                 & 8b                & 7                      & 21.10M              & 0.10 / 16                                                                \\
             & ResNet-14      & CIFAR10          & 8b                 & 8b                & 13                     & 35.30M               & -0.32 / 20                                                              \\
             & ResNet-50      & CIFAR10          & 8b                 & 8b                & 49                     & 0.11G             & -0.15 / 17                                                              \\ \hline
\multirow{3}{*}{\cite{Hammad2021:controller}}     & VGG-19         & ImageNet         & 16b                 & 16b               & 16                     & 19.64G            & 0.47 / 81                                                               \\
             & Xception       & ImageNet         & 16b                 & 16b               & 36                     & 8.40G            & 0.90 / 81                                                                \\
             & DenseNet201    & ImageNet         & 16b                 & 16b               & 200                    & 3.35G             & 1.14 / 85                                                               \\ \hline
\multirow{4}{*}{\cite{Riaz2020:CaxCNN}}     & LeNet        & MNIST            & 8b                 & 8b                & 2                      & 0.34M               & 0.03 / -\tnote{1}                                                              \\
             & CIFAR10        & CIFAR10          & 8b                 & 8b                & -                      & -                 & 0.62 / -\tnote{1}                                                               \\
             & AlexNet        & ImageNet         & 8b                 & 8b                & 5                      & 1.13G               & 0.02 / -\tnote{1}                                                               \\
             & VGG-16         & ImageNet         & 8b                 & 8b                & 13                     & 15.48G              & 4.80 / -\tnote{1}                                                              \\ \hline
{\cite{Guo2020:DoubleModeMult}}     & VGG-16         & ImageNet         & 8b / 16b           & 16b               & 13                     & 15.48G              & 3.00 / 37                                                                     \\ \hline
\multirow{12}{*}{\cite{Zervakis2021:ControlVar}$^\dagger$}     & GoogleNet      & CIFAR10          & 8b                 & 8b                & 22                     & 0.76G                & -0.16 / 23                                                               \\
             & GoogleNet      & CIFAR100         & 8b                 & 8b                & 22                     & 0.76G                & 0.05 / 23                                                                \\
             & ResNet-44      & CIFAR10          & 8b                 & 8b                & 43                     & 97.80M               & 0.03 / 23                                                                \\
             & ResNet-44      & CIFAR100         & 8b                 & 8b                & 43                     &97.80M               & 0.77 / 23                                                                \\
             & shufflenet     & CIFAR10          & 8b                 & 8b                & 3                      & 80.85M              & -0.48 / 35                                                               \\
             & shufflenet     & CIFAR100         & 8b                 & 8b                & 3                      & 80.85M              & 0.20 / 23                                                                 \\
             & VGG-13         & CIFAR10          & 8b                 & 8b                & 10                     & 0.23G             & -0.30 / 35                                                                \\
             & VGG-13         & CIFAR100         & 8b                 & 8b                & 10                     & 0.23G             & 0.89 / 23                                                                \\
             & VGG-16         & CIFAR10          & 8b                 & 8b                & 13                     & 0.15G               & 0.38 / 35                                                                \\
             & VGG-16         & CIFAR100         & 8b                 & 8b                & 13                     & 0.15G               & 0.03 / 35                                                                \\
             & ResNet-56      & CIFAR10          & 8b                 & 8b                & 55                     & 0.13G             & 0.49 / 23                                                                \\
             & ResNet-56      & CIFAR100         & 8b                 & 8b                & 55                     & 0.13G             & -0.34 / 23                                                               \\ \hline 
\multirow{7}{*}{\cite{Tasoulas2020:weightoriented}$^\dagger$}      & ResNet-20      & LISA             & 8b                 & 8b                & 21                     & 40.80M             & 0.50 / 20                                                                 \\
             & ResNet-32      & GTSRB            & 8b                 & 8b                & 33                     & 69.10M               & 0.50 / 15                                                                   \\
             & ResNet-44      & LISA             & 8b                 & 8b                & 45                     & 97.40M               & 0.50 / 20                                                                 \\
             & ResNet-56      & LISA             & 8b                 & 8b                & 57                     & 0.13G             & 0.50 / 22                                                                 \\
             & MobileNet-V2   & CIFAR100         & 8b                 & 8b                & 35                     & 82.10M              & 2.00 / 19                                                                   \\
             & VGG-11         & CIFAR10          & 8b                 & 8b                & 8                      & 153M              & 1.00 / 19                                                                   \\
             & VGG-13         & CIFAR100         & 8b                 & 8b                & 10                     & 0.23G              & 0.50 / 19                                                                 \\ \hline
{\cite{Zervakis:ACCESS2020}}           & ResNet-8       & CIFAR10          & 8b                 & 8b                & 7                      & 21.10M              & 0.50 / 19                                                                   \\ \hline
                                                  
\end{tabular}
\begin{tablenotes}
\item[1] \cite{Riaz2020:CaxCNN} reports only area benefits in terms of BELs usage from 45\% up to 97\%
\item[*] \blue{Retraining/Fine-tuning is used  (see Section~\ref{subsec:retraining})}
\item[$^\dagger$] \blue{Statistical error compensation technique is used (see Section~\ref{subsec:error})}
\end{tablenotes}
\end{threeparttable}%
\end{table}

Table~\ref{tab:axun_1} reports the respective results for the Approximate Multipliers/Adders family (subcategory of Approximate Units).
This is the largest approximation family, comprising the most works.
The latter can be explain by the vast research activities that focused on approximate multiplication and addition circuits since the introduction of approximate computing~\cite{ApproxCircuitsSurvey}.
As shown in Table~\ref{tab:axun_1}, the approximation techniques of this family mainly target low precision implementations (i.e., $8$-bit mostly) and examine a variety of datasets and DNNs (from simple to more complex ones).
As in the previous approximation categories, when considering less complex benchmarks, very high energy savings are achieved, combined with minimal accuracy loss.
For example, for LeNet-5 on MNIST,~\cite{Mrazek2016:CGPevo} delivered $77$\% energy reduction and only $0.09$\% accuracy loss.
Similarly, using $16$-bit precision, \cite{Hammad2021:controller} attained $81$\% energy savings and $0.47$\% accuracy loss for VGG-19 on ImageNet.
In more complex evaluations, the obtained energy gains are still significant albeit being decreased.
Remarkably, using $8$-bit the dynamic weight-oriented approximation of~\cite{Tasoulas2020:weightoriented} achieved $19$\% energy reduction and $2$\% accuracy loss for MobileNet on CIFAR100 while the curable control variate approximation of~\cite{Zervakis2021:ControlVar} delivered $35$\% energy savings and only $0.03$\% accuracy loss for VGG-16 on CIFAR100.
Compared to the $8$-bit baseline,~\cite{Mrazek2020:Libraries} used an approximate $8\times4$ multiplier ($8$ bits for the activations and $4$ bits for the weights) and achieved $70$\% energy reduction and $1.21$\% accuracy loss for ResNet-26 on CIFAR10.
Nevertheless, compared to the accurate $8\times4$ multiplier, these values translate to $15$\% energy reduction and $0.08$\% accuracy loss.

\begin{figure}[t!]
   \centering
    \includegraphics[width=1\textwidth]{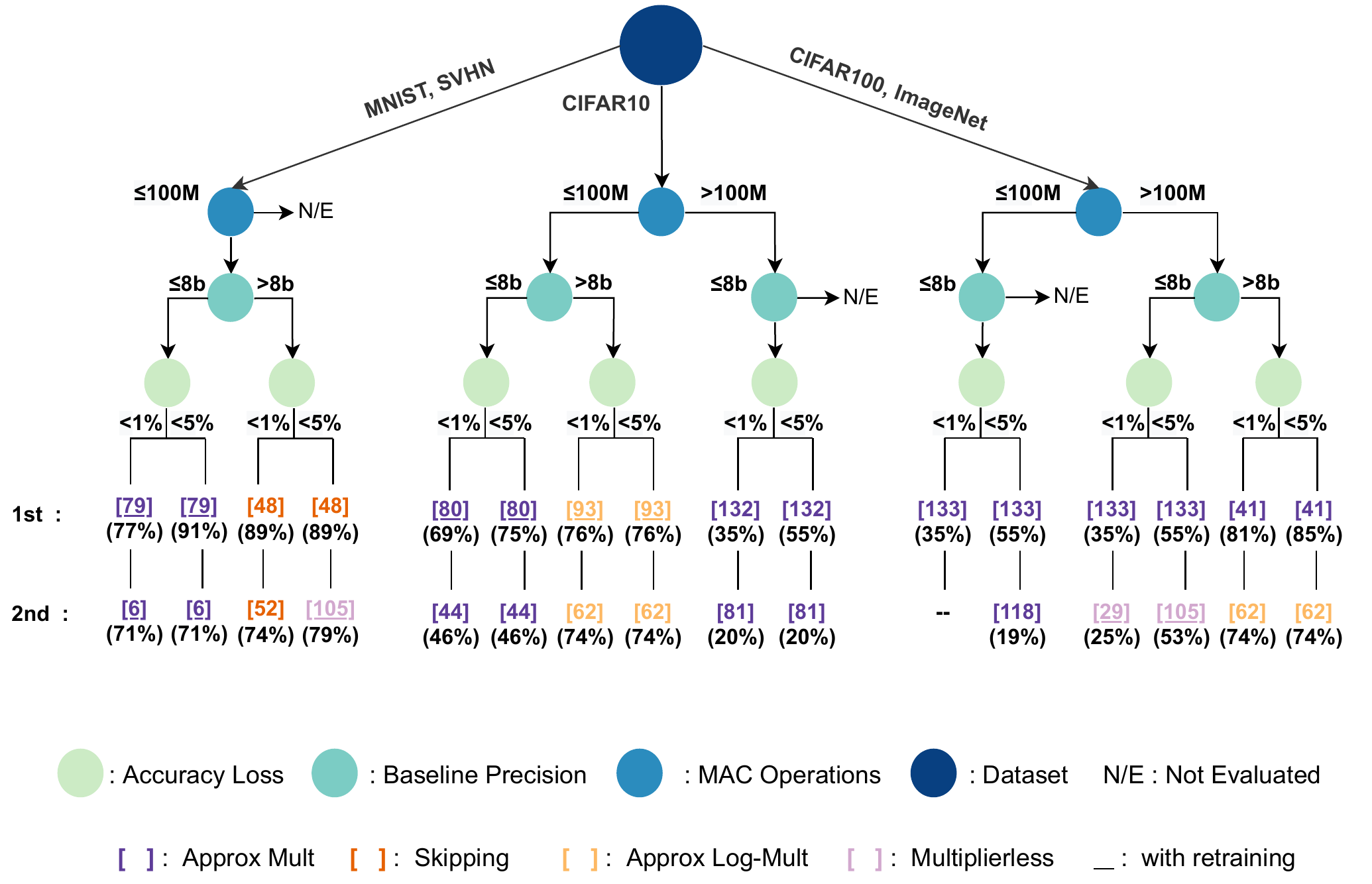}
\caption{
Classification of the optimal hardware approximation techniques with respect to the complexity of the performed evaluation, i.e., dataset, DNN size, precision, and an accuracy loss threshold ($1$\% and $5$\%).
The leafs present the corresponding optimal approximate technique (i.e., highest energy reduction) as well as the respective runner up technique.
Below each technique the attained energy reduction is reported.
The approximate color-coded with respect to the approximation family (see Fig.~\ref{fig:Clustering}) that they belong.
Underlined techniques require retraining.
}
    \label{fig:tree}
\end{figure}

Finally, the above analysis (Tables~\ref{tab:compred}-\ref{tab:axun_1}) is summarized in Fig.~\ref{fig:tree}.
To generate Fig.~\ref{fig:tree}, we identified the most widely used datasets as well as the most common network sizes ($\leq 100$M or $>100M$ MAC operations\footnote{As an example,  ResNet-44 on CIFAR10 requires 97.8M MAC operations.}) and we set two precision levels, i.e., low precision ($\leq 8$ bits) and high precision ($> 8$ bits).
Then, we created a decision tree that helps the reader to identify the optimal approximation technique/family with respect to the complexity of the evaluation (i.e., dataset , size of the network, and precision) and the desired accuracy loss constraint.
%The number of higher than 100M MAC operations corresponds to networks bigger than ResNet-44. 
For the accuracy loss we considered two thresholds, i.e., small accuracy loss (less than $1$\%) and moderate accuracy loss (less than $5$\%).
The leafs represent the two approximation techniques (highest and runner-up) that achieve the highest energy reduction in each case.
In addition, the respective energy reducton is also reported below the corresponding technique.
The techniques are color-coded with respect to the approximation family that they belong and underlined techniques require DNN retraining to achieve the respective accuracy loss threshold.
Note that some tree branches are empty since the respective cases haven't been considered in the evaluation of the examined approximate DNN works.
For example, considering the MNIST/SVHN datasets, only small DNNs are evaluated since they can achieve almost perfect accuracy.
%This is mainly because works that evaluate their technique on easy and medium difficulty datasets avoid complex networks with many parameters and consequently many MAC operations, as small ones are enough to predict correctly the tested image.
As shown in Fig.~\ref{fig:tree}, when high precision is used (mainly $32$-bit) the energy savings are maintained and remain very high (more than $74$\%) in all cases.
However, when low precision is examined (mainly $8$-bit) the energy savings mainly drop as the complexity of the evaluation increases.

Overall, the Approximate Multipliers/Adders family dominates Fig.~\ref{fig:tree} but this might also be subject to the fact that the Approximate Multipliers/Adders is the largest examined approximation family.
Nonetheless, we can observe a considerable diversity in the tree's leafs with respect to the approximation family.
\blue{Specifically for the Approximate Units category (Section~\ref{subsec:axunits}), as shown in Fig.~\ref{fig:tree}, techniques from all the three families (i.e., Multiplierless, Approximate Log-Multipliers, Approximate Multipliers/Adders) appear among the optimal solutions.\label{commentR3C12}
When high precision is required then the Approximate Log-Multipliers constitute mainly the best solutions.
On the other hand, considering the most challenging evaluation (i.e., complex dataset and low bitwidth) then the Approximate Multipliers/Adders and Multiplierless families prevail.
In is noteworthy that Multiplierless is represented by two different works (\cite{Sarwar2018:Alphabet} and~\cite{Faraone2020:AddNet}) that both require retraining however.
On the other hand, the Approximate Multipliers/Adders is represented by one work (i.e.,~\cite{Zervakis2021:ControlVar}) that employs a statistical error compensation method (see Section~\ref{subsec:error}).}
\blue{It is noteworthy that although the impact of the error compensation techniques is not always comprehensively analyzed in the respective works, from Fig.~\ref{fig:tree} we can deduce that such techniques are mandatory to achieve high energy savings combined with low accuracy loss.
}\label{commentR2C3b}

Moreover, we observe, in Fig.~\ref{fig:tree}, that the approximation techniques that require or not retraining are quite balanced.
Out of all the techniques reported in the leafs of Fig.~\ref{fig:tree}, ~\cite{Mrazek2016:CGPevo,Mrazek2020:Libraries,TwoStageLog:2021,Ansari2020:ImprovEvo,Sarwar2018:Alphabet,Faraone2020:AddNet} require retraining while~\cite{Hemmat2020a:AirNN,Zervakis2021:ControlVar,Hammad2021:controller,Huan2016:NearZero,Hanif2019:CANN,Kim2019:mitchell,Mrazek2019:ALWANN,Tasoulas2020:weightoriented} do not.
Hence, although significant research focused on approximation-aware retraining, in the challenging evaluations (e.g., ImageNet in Fig.~\ref{fig:tree}), the optimal techniques do not apply retraining.
The latter could be explained by fact that retraining for ImageNet is very time consuming.
This further highlights the need for more efficient approximation-aware retraining or for techniques that apply curable approximations without retraining.
Finally, hardware approximation for DNNs can deliver significant energy savings even when considering complex scenarios and low accuracy loss constraints.
However, by highlighting the high difference in energy savings between the ``easy'' and ``complex'' evaluation scenarios, it is derived that significant research is still required in more challenging benchmarks.

\section{Not Just Energy Efficiency}\label{sec:other}

In the previous sections, we analyzed the impact of approximate computing on the energy efficiency and accuracy of DNN accelerators, demonstrating that significant energy gains are achieved for a minimal accuracy loss.
In this section, our analysis goes beyond the energy efficiency of DNNs.
Recent research has shown that approximate computing principles enable designers to overcome degradation effects (reliability-aware approximation) and security weaknesses (defensive approximation) of DNN accelerators.
%Nevertheless, recent research has shown that principles from AC can enable designers to also overcome some other degredation effects, such as temperature~\cite{temp_approx:Henk2018}, aging~\cite{aging_approx:Henk2020} and security weakness~\cite{Gonzalez2019:security} of DNN applications, that without a carefull control each of them can lead to catastrophic failures.
%The purpose of this section is to present how approximations can enhance two fields in which research activity is still limited, but critical.

\subsection{Reliability-Aware Approximation}

In contrast to traditional thermal management approaches~\cite{managementSurvey:Henk}, Amrouch et al.~\cite{Amrouch2020:NpuThermal} employed approximate computing as a solution.
Through precise chip modeling and multi-physics simulations using Ansys,~\cite{Amrouch2020:NpuThermal} demonstrated that DNN accelerators are subject to excessive power density and elevated on-chip temperatures.
% Through precise chip modeling and multi-physics simulations using Ansys, Amrouch et al.~\cite{Amrouch2020:NpuThermal} demonstrated that DNN accelerators are subject to excessive power density and elevated on-chip temperatures.
\cite{Amrouch2020:NpuThermal} proposed the first hybrid thermal management for DNN accelerators that employs runtime approximation as a cooling mechanism.
Dynamic precision scaling through clock gating and low bitwidth quantization is applied in~\cite{Amrouch2020:NpuThermal}.
As a result, at the cost of some accuracy loss, reduced switching activity and thus lower power and power density are achieved.
Hence, as \cite{Amrouch2020:NpuThermal} demonstrated, for the same cooling cost, precision scaling can reduce the power and thus the temperature.
Similarly, the power gain of precision scaling can be traded to increase both the frequency and the cooling cost and achieve higher performance for same temperature and total power consumption.
It is noteworthy that for $85$\textcelsius{} temperature constraint, precision scaling boosted the efficiency (throughput/Joule) of the DNN accelerator by $1.5$x~\cite{Amrouch2020:NpuThermal}.

The very high utilization of the DNN accelerator's MAC units exposes the underlying transistors to continuous stress with very little time for relaxation and recovery~\cite{salamin2021:AgingAware}.
As a result, transistors age faster.
The presence of excessive temperatures~\cite{Amrouch2020:NpuThermal} exacerbates further the problem as the majority of mechanisms behind transistor aging exponentially depend on the operating temperature~\cite{salamin2021:AgingAware}.
Salamin et al.~\cite{salamin2021:AgingAware} proposed a circuit aging aware approximation framework that applies a graceful-approximation technique to suppress, over time, aging effects in DNN accelerators.
Through aging-aware cell libraries,~\cite{salamin2021:AgingAware} analyzed the delay of MAC units at varying aging levels.
Exploiting that lower bit-width inputs activate shorter paths and thus the circuit can operate faster,~\cite{salamin2021:AgingAware} performed static time analysis to identify the maximum precision for each MAC input so that no aging induced timing errors occur.
The obtained precision was used to quantize the weights and activations of the NN at runtime.
A library of quantization methods~\cite{Krishnamoorthi2018:QuantWhitePap,Jacob2018:Quant,Banner2019:ACIQ} was used to achieve the highest accuracy.
\cite{salamin2021:AgingAware} eliminated the aging-induced timing gurdbands, boosting the throughput by $23$\%, and delivered a progressive accuracy degradation over time.
At $10$ years aging, the accuracy loss was only $2.96$\% on average~\cite{salamin2021:AgingAware}.

%AS FAR AS I UNDERSTOOD, DNN LIFE IS NOT APPROXIMATE
%In~\cite{hanif2021dnnlife}, they demonstrated that an optimal duty-cycle can balance aging of weight memory cells of a given DNN hardware that are used during the inference process.
%Leveraging this fact, they proposed a framework including an aging controller and a data encoding scheme, which accounts for varied DNN workloads to adapt so that they balance the duty-cycle in each of the on-chip weight memory cell.
%The role of aging controller is to generate the proper encoding information required so that duty-cycle is balanced.
%Examining different DNN data representation formats, this framework enables efficient aging mitigation of weight memory, but with a small delay, area and energy overhead.

\subsection{Defensive Approximations}

\begin{figure}[t]
 \centering
\resizebox{0.9\textwidth}{!}{\includegraphics[]{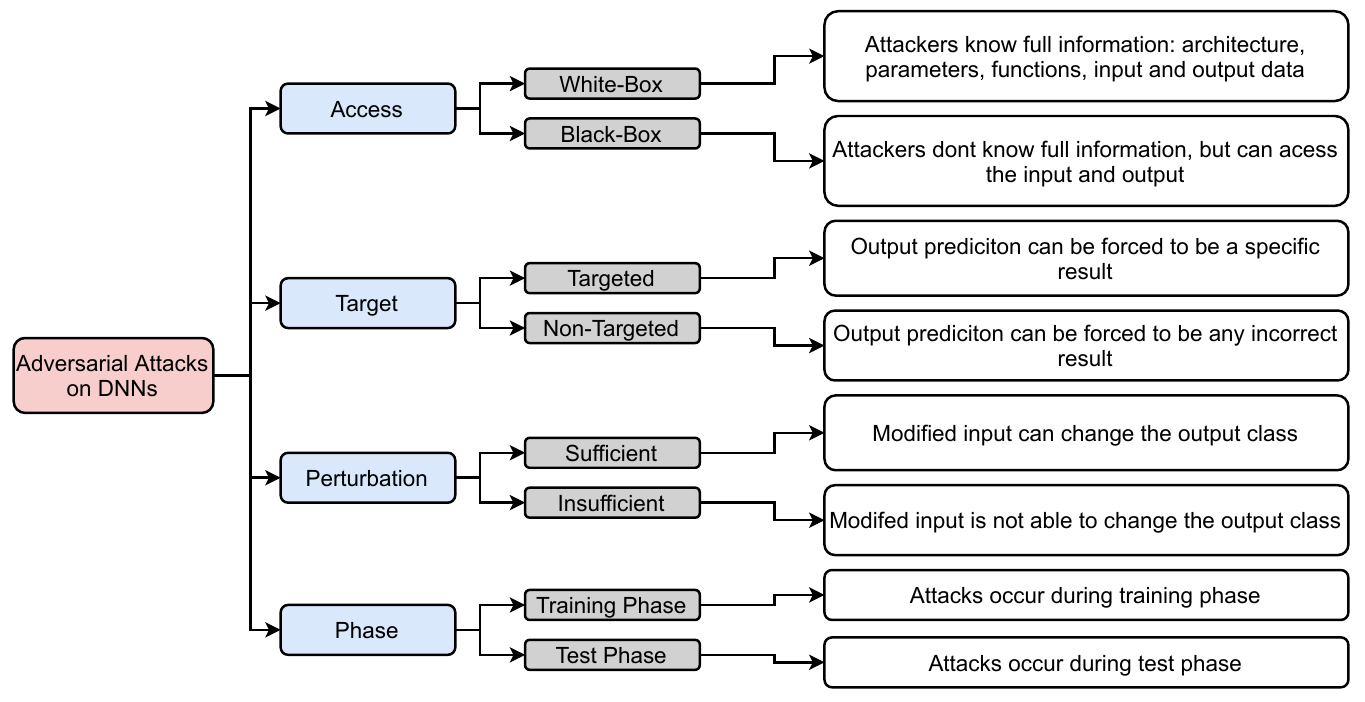}}%
\centering
\caption{
Classification of DNN Adversarial Attacks.
}
\label{fig:attacks}%
\end{figure}

Recent research showed that DNNs have intrinsic security weaknesses and are susceptible to adversarial attacks~\cite{Gonzalez2019:security, Shafique2020:security}.
% , \orange{causing the target to deviate from its expected behaviour during inference.}
Approximate computing has emerged as a means for making DNN models more robust against such attacks, while maintaining its effective trade-offs, i.e., high energy savings for a small accuracy loss.
The scope of an adversarial attack, \yellow{in the case of vision tasks}, is to introduce a noise in the input image to create a spiteful sample, which is misclassified by the DNN. 
Two categories of adversarial attacks that have preoccupied defensive approximations, regarding the knowledge of attackers, are White-Box attacks (when attackers know the training data, model parameters, and model architecture) and Black-Box attacks (when they don't know model information)~\cite{BlackBox}.
The rest of them can be categorized according to different properties, e.g., the target of attack, the kind of perturbation and the phase in which an attack occurs.
These attributes are summarized and explained in Fig.~\ref{fig:attacks}.

Most of the proposed methods that used approximate computing as defense aim at increasing the generalization of DNNs, as observations show they perform better against various types of attacks.
It has been also noticed that low-precision models exhibit, in general, higher adversarial accuracy compared to the full-precision models with identical network structures~\cite{AttackingBNN:2019,galloway2018AttackingBNN}.
This could be explained due to quantization effect that enhances the amount of non-linearity, which prevents small changes in the input from drastically altering the output and forcing a misclassification.

%\blue{
Guesmi et al.~\cite{guesmi2020defensive}, in order to handle such threats, proposed an approximate CNN implementation, where the exact multiplier replaced with an approximate FP multiplier that injects data-dependent noise in convolution operations.
This approximate mantissa multiplier induced an error, which was propagated through the whole model.
It was observed that this error could further the difference between the first class and the ``runner-up'' and help the classifier to generalize and enhance its confidence.
Experimental evaluation over LeNet-5 trained on MNIST showed that, for a negligible accuracy loss, the proposed defensive approximate scheme made the model $87.5\%$ more robust against Black-Box attacks than the conventional CNN, with $50\%$ power and $67\%$ area reduction.

% Fine-Pruning~\cite{FinePruning2018} focuses on eliminating the effect of the backdoor (training-time attacks) by finding the redundant connections in DNNs that do not significantly contribute for the accuracy of the clean data, and removing them.
% The operation of pruning operates in three phases in which firstly, neurons activated by neither clean nor backdoored inputs are pruned, secondly, neurons activated by the backdoor but not by clean inputs and thirdly, neurons activated by clean inputs.
% The process stops when the accuracy drops below a pre-defined threshold.
% Once the model architecture and training data have been determined, fine-pruning can achieve much lower computational cost than the conventional model.
% Substantially, it combines the defensive strengths of both pruning and fine-tuning and nullifies backdoor attacks to 0\% in some cases with only a small degradation in accuracy (<1\%). \red{I Don't remember, we keep Fine-Pruning or not?}

%\blue{

% \red{1. HOW DOES IT FIND REDUNDANT CONNECTIONS? 2. ELABORATE WHICH ARE THE DEFENSIVE STRENGTHS OF BOTH PRUNING AND FINE-TUNING. 3. IS THIS HARDWARE APPROXIMATION?}

A quantization-based defense, which exploits low precision implementations, was proposed in~\cite{QuantDefense}.
First,~\cite{QuantDefense} selects a number of quantization levels based on the application's resilience to errors and perturbations.
Next, an additional layer is added at the input of the network that has one-to-one relation with the input pixels.
This relation with the rest quantization scheme supports different configurations based on whether training is needed or not.
The idea of this proposed defense is based on the observation that when the input of a CNN is quantized, the confidence of a clean image's prediction remains almost the same.
On the contrary, the confidence of an incorrectly classification of a perturbed image is decreased.
Evaluating the proposed scheme under different white-box and black-box settings showed an increase in the classification accuracy of perturbed images by up to 50\% and 96\% for CIFAR10 and MNIST datasets, respectively.

\section{Conclusions, Challenges, and Perspective}

In this article, hardware approximation techniques for DNNs are reviewed,
characterized, classified and evaluated.
Moreover, we provide a comprehensive and analysis of error metrics and error mitigation approaches  specific for DNN approximations in order to provide an in depth analysis of the studied field.
Note that, in addition to the traditional exploitation of Approximate Computing for energy reduction, we present how approximate computing can be employed in DNNs to address reliability and security concerns. 
Our analysis clustered the hardware DNN approximation techniques in three categories: Precision Scaling, Computation Reduction, and Approximate Units.

Precision scaling is the most widely used method and already adopted by most commercial DNN accelerators.
Advancements quantization aware (re-)training methods have led to minimal accuracy loss even with $4$-bit or $2$-bit inference.
%At the cost of reduced throughput, high precision arithmetic computations can be obtained enabling full accuracy inference.
Nevertheless, note that quantization aware training can be very time consuming and post-training quantization approaches are efficient for $8$-bit inference -- that is considered mainstream today -- and with some limitations might enable $4$-bit inference.

The Computation Reduction approximation is demonstrated to deliver very high energy reduction for minimal accuracy loss.
However, this approximate category mainly examines $32$-bit inference and the energy gains dropped significantly when considering more challenging evaluations such as $8$-bit inference and/or ImageNet.
As a result, to obtain conclusive results regarding the Computation Reduction a more in depth comprehensive analysis is required either with respect to more challenging evaluation scenarios or to NNs that indeed require high precision inference.

% After Precision Scaling, the Approximate Units category has attracted the highest research interest.
% Typically, Approximate Units is combined with low precision (mainly $8$-bit inference).
% Although for small DNNs the Approximate Multipliers/Adders family delivers immense energy reduction with negligible accuracy loss, overall evaluations on the state of the art ImageNet datasets are still limited.
% Moreover, although some works aimed at evaluating the trade-off between low precision ($8$-bit and below) and approximate units in DNN inference and showed that a combination of the two outperforms the isolated application of very low precision, this correlation is not comprehensively analyzed yet.
% Finally, an inherent limitation in this category is that many techniques require retraining to recover the accuracy loss.
% In contrast to quantization-aware training that can run efficiently on CPUs and GPUs, retraining with approximate units requires hardware emulation that can even become infeasible in complex DNNs.
% Hence, despite the research efforts made in this direction, approximation-aware DNN retraining is still at its infancy.
% To avoid retraining, curable approximation and/or statistical error compensation employment appear to be very promising solutions but are still understudied.
After Precision Scaling, the Approximate Units category has attracted the highest research interest.
Typically, Approximate Units is combined with low precision (mainly $8$-bit inference).
Again, despite the high energy gains reported in many cases, a more comprehensive and challenging evaluation is required.
Although the results seem promising, evaluations on the state of the art ImageNet datasets are still limited.
Still, it is noteworthy that for small DNNs the Approximate Multipliers/Adders family delivers immense energy reduction with negligible accuracy loss.
The latter appears ideal for IoT devices that need to run sophisticated DNN-based services.
Moreover, although some works aimed at evaluating the trade-off between low precision ($8$-bit and below) and approximate units in DNN inference and showed that a combination of the two outperforms the isolated application of very low precision, this correlation is not comprehensively analyzed yet.
Finally, note that an inherent limitation is this category that many techniques require retraining to recover the accuracy loss.
In contrast to quantization-aware training that can run efficiently on CPUs and GPUs, retraining with approximate units requires hardware emulation that can even become infeasible in complex DNNs.
%Hence, despite the research efforts made in this direction, approximation-aware DNN retraining is still at its infancy.
To avoid retraining, curable approximation and/or statistical error compensation methods appear to be very promising solutions but are still understudied.

Hardware approximation for DNNs has shown remarkable advancements over the past years moving from simple DNNs to very complex ones.
Although, Approximate Computing has demonstrated a great potential through some impressive results, still, significant innovation is required to enable hardware approximation to be actively adopted in the design of complex DNN accelerators.
% for DNN circuits is not yet fully mature, it 
Finally, a crucial but not well addressed topic is the relation between approximate computing and the standardization of ML-based systems in safety critical applications.\label{commentR1C12}
Safety standards for systems ML-based are yet to be formalized~\cite{tambon2021certify} and as a result, the impact of approximate computing on the system’s certification remains unclear.
The examined hardware approximation techniques are deterministic and are not expected to impact the certification process.
On the other hand, despite the high inference accuracy achieved by these approximations, the ML system will not work as it was trained to do (due to the induced approximation during inference).
Thus, this might hinder the certification of approximate ML-based systems in safety-critical scenarios.
Nevertheless, such issues could be solved through approximation-aware retraining.
As discussed in Section~\ref{sec:other}, approximate computing enhances the reliability and robustness of DNN accelerators and thus might ease the certification of the system.
Overall, a deep investigation and analysis is required.

\begin{acks}
This work is partially supported by the German Research Foundation (DFG) through the project ``ACCROSS: Approximate Computing aCROss the System Stack'' HE 2343/16-1.
\end{acks}


\end{document}